\documentclass{aa}
\usepackage{graphicx}
\usepackage{longtable}
\usepackage{txfonts}
\usepackage[authoryear]{natbib}
\bibpunct{(}{)}{;}{a}{}{,}

\begin{document}

\title{Radio continuum and near-infrared study of the \object{MGRO~J2019+37}
region}
\authorrunning{Paredes et al.}
\titlerunning{Radio and NIR study of MGRO~J2019+37}

\author{
J.M. Paredes\inst{1}
\and J.Mart\'{\i}\inst{2,3}
\and C.H. Ishwara-Chandra\inst{4}
\and J.R. S\'anchez-Sutil\inst{3}
\and A.J. Mu\~noz-Arjonilla\inst{2,3}
\and J. Mold\'on\inst{1}
\and M. Peracaula\inst{5}
\and P.L. Luque-Escamilla\inst{3}
\and V. Zabalza\inst{1}
\and V. Bosch-Ramon\inst{6}
\and P. Bordas\inst{1}
\and G.E. Romero\inst{7,8}\thanks{Member of CONICET}
\and M. Rib\'o\inst{1}
}


\institute{
Departament d'Astronomia i Meteorologia and Institut de Ci\`encies del Cosmos (ICC), Universitat de Barcelona (UB/IEEC), Mart\'{\i} i Franqu\`es 1, 08028 Barcelona, Spain \\
\email{jmparedes@ub.edu, jmoldon@am.ub.es, vzabalza@am.ub.es, pbordas@am.ub.es, mribo@am.ub.es}
\and Departamento de F\'{\i}sica, EPS, Universidad de Ja\'en, Campus Las Lagunillas s/n, Edif. A3, 23071 Ja\'en, Spain \\
\email{jmarti@ujaen.es, ajmunoz@ujaen.es} 
\and Grupo de Investigaci\'on FQM-322, Universidad de Ja\'en, Campus Las Lagunillas s/n, Edif. A3, 23071 Ja\'en, Spain \\
\email{jrssutil@ujaen.es, peter@ujaen.es}
\and NCRA, TIFR, Post Bag 3, Ganeshkhind, Pune-411 007, India \\
\email{ishwar@ncra.tifr.res.in}
\and Institut d'Inform\`atica i Aplicacions, Universitat de Girona, Girona, Spain \\
\email{marta.peracaula@udg.edu}
\and Max Planck Institut f\"ur Kernphysik, Saupfercheckweg 1, Heidelberg 69117, Germany \\
\email{vbosch@mpi-hd.mpg.de}
\and Instituto Argentino de Radioastronom\'{\i}a (CCT La Plata, CONICET), C.C.5,
(1894) Villa Elisa, Buenos Aires, Argentina\\
\email{romero@iar.unlp.edu.ar}
\and Facultad de Ciencias Astron\'omicas y Geof\'{\i}sicas, UNLP, 
Paseo del Bosque, 1900 La Plata, Argentina \\
\email{romero@fcaglp.unlp.edu.ar}
}

\date{Received / Accepted}

\abstract
{\object{MGRO~J2019+37} is an unidentified extended source of very high energy
gamma-rays originally reported by the Milagro Collaboration as the brightest
TeV source in the Cygnus region. Its extended emission could be powered by either a single or several sources. The GeV pulsar \object{AGL~J2020.5+3653},
discovered by AGILE and associated with \object{PSR~J2021+3651}, could
contribute to the emission from \object{MGRO~J2019+37}.}
{Our aim is to identify radio and near-infrared sources in the field of the
extended TeV source \object{MGRO~J2019+37}, and study potential counterparts to
explain its emission.}
{We surveyed a region of about 6 square degrees with the Giant Metrewave
Radio Telescope (GMRT) at the frequency 610~MHz. We also observed the central
square degree of this survey in the near-infrared $K_{\rm s}$-band using the
3.5~m telescope in Calar Alto. Archival X-ray observations of some specific
fields are included. VLBI observations of an interesting radio source were 
performed. We explored possible scenarios to produce the multi-TeV
emission from \object{MGRO~J2019+37} and studied which of the sources could be 
the main particle accelerator.}
{We present a catalogue of 362 radio sources detected with the GMRT in the
field of \object{MGRO~J2019+37}, and the results of a cross-correlation of this
catalog with one obtained at near-infrared wavelengths, which contains
$\sim 3\times10^{5}$ sources, as well as with available X-ray observations of
the region. Some peculiar sources inside the $\sim$1\degr\ uncertainty region
of the TeV emission from \object{MGRO~J2019+37} are discussed in detail,
including the pulsar \object{PSR~J2021+3651} and its pulsar wind nebula
\object{PWN~G75.2+0.1}, two new radio-jet sources, the \ion{H}{ii} region
\object{Sh~2-104} containing two star clusters, and the radio source
\object{NVSS~J202032+363158}. We also find that the hadronic scenario is the
most likely in case of a single accelerator, and discuss the possible
contribution from the sources mentioned above.}
{Although the radio and GeV pulsar \object{PSR~J2021+3651} /
\object{AGL~J2020.5+3653} and its associated pulsar wind nebula
\object{PWN~G75.2+0.1} can contribute to the emission from
\object{MGRO~J2019+37}, extrapolation of the GeV spectrum does not explain the
detected multi-TeV flux. Other sources discussed here could contribute to the
emission of the Milagro source.}


\maketitle

\section{Introduction} \label{introduction}

The Galactic very-high-energy (VHE) $\gamma$-ray sources discovered by the 
latest generation of Cherenkov observatories (H.E.S.S., MAGIC,
Milagro) are currently an actively studied topic in modern high-energy astrophysics. Among
the $\sim$75 detected sources, nearly one third remain yet as unidentified. A
significant number of them have extended morphologies on 0.1--1\degr\ scales in
the TeV energy band, ensuring that the identification of counterparts at lower
energies is a very difficult task. The most representative of this new population
of Galactic sources is \object{TeV~J2032+4130}, inside whose error box both
compact and extended radio sources on arcsecond scales were found
\citep{paredes07}. {\it XMM-Newton} observations of this source also detected 
faint extended X-ray emission \citep{horns07a}.

An addition to the population of extended, unidentified TeV sources was 
reported by the Milagro collaboration, following the discovery in the Cygnus region
of the most extended TeV source known so far
\citep{abdo07_cygnus,abdo07_survey}. The TeV emission from this area covers
several square degrees and includes diffuse emission and at least one new source,
\object{MGRO~J2019+37}, located to within an accuracy of $\pm$0.4\degr. 
After the
Crab Nebula, \object{MGRO~J2019+37} is the strongest source detected by Milagro. The Tibet AS-$\gamma$ experiment confirmed the detection of
this source by measuring a 5.8$\sigma$ signal compatible with the position of
\object{MGRO~J2019+37} \citep{amenomori08}. On the other hand, VERITAS inferred  an upper limit that is compatible with the Milagro detection for a hard-spectrum extended source \citep{kieda08}.

\begin{figure*}
\center
\resizebox{\hsize}{!}{\includegraphics[angle=0]{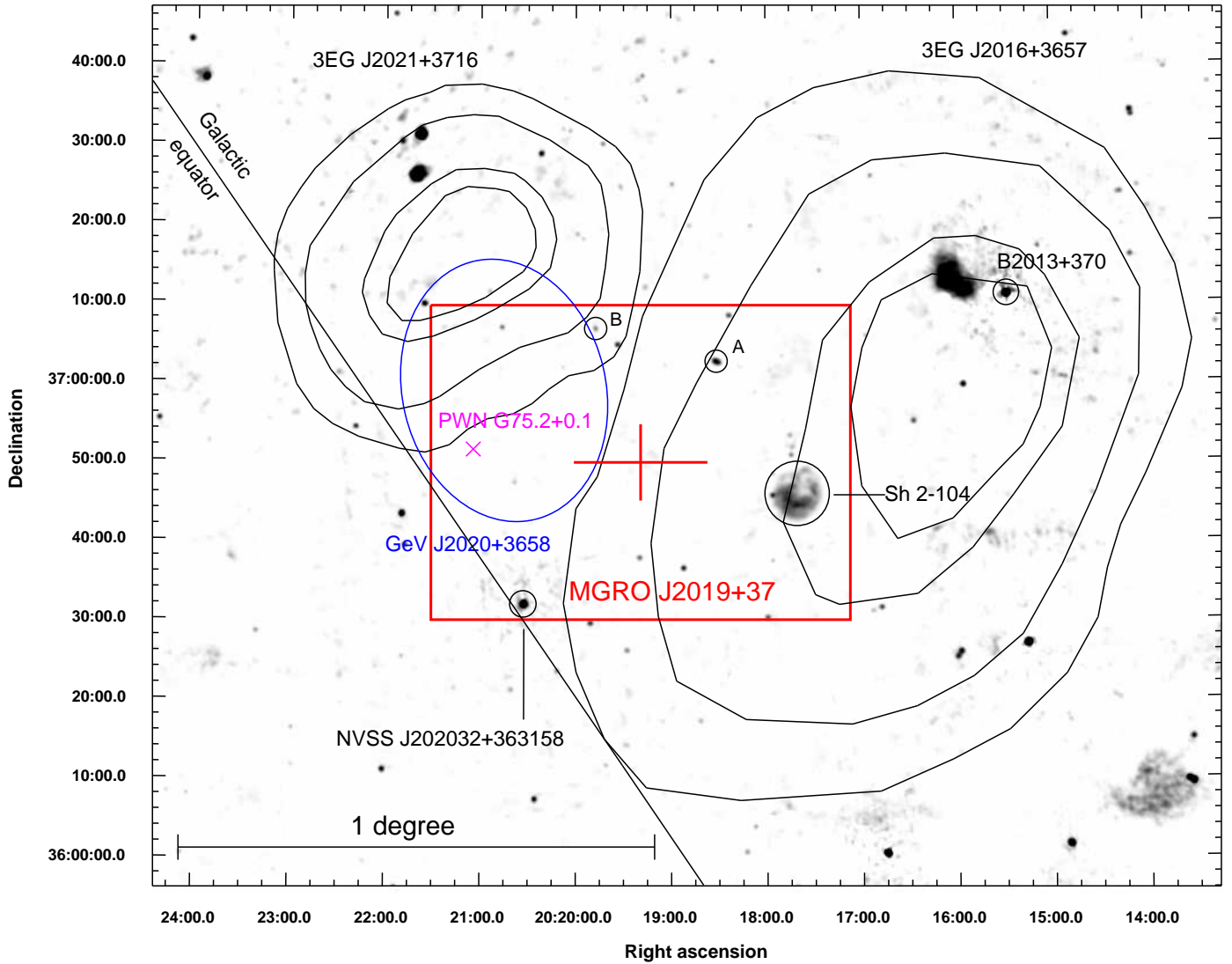}}
\caption{Radio map obtained with the GMRT at 610~MHz (greyscale) convolved with
a circular restoring beam of 30\arcsec. The red cross and box indicate the
center of gravity and its positional uncertainty including statistic and
systematic errors of the TeV emission from the source \object{MGRO~J2019+37}
\citep{abdo07_survey}. The conspicuous radio sources located inside this box
correspond to the extended \ion{H}{ii} region \object{Sh~2-104}, a bright
compact radio source also detected with the VLA as
\object{NVSS~J202032+363158}, and two newly discovered jet-like sources (A and
B). The position probability contours (50\%, 68\%, 95\%, and
99\%, from inside to outside) of the Third EGRET catalogue sources
\object{3EG~J2021+3716} and \object{3EG~J2016+3657} \citep{hartman99}, as well
as the GeV source \object{GeV~J2020+3658} (blue ellipse) \citep{lamb97} are superimposed. The
magenta cross indicates the position of the pulsar wind nebula
\object{PWN~G75.2+0.1}. The blazar \object{B2013+370} within
\object{3EG~J2016+3657} is also labeled.}
\label{fig:mapa}
\end{figure*}

The origin of all these types of emission and their association with astrophysical sources
is unclear. Although a possible connection with the anisotropy of Galactic
cosmic rays was proposed \citep{amenomori06}, the TeV $\gamma$-ray flux 
measured at 12~TeV from the diffuse emission of the Cygnus region (after
excluding \object{MGRO~J2019+37}), exceeds that predicted by a conventional
model of cosmic ray production and propagation \citep{abdo07_cygnus}. This
strongly infers the existence of hard-spectrum cosmic-ray sources and/or
other types of TeV $\gamma$-ray sources in the region. It is unclear whether
the emission originates in either a single extended source or a combination
of several point sources. \object{MGRO~J2019+37} is positionally coincident
with the EGRET sources \object{3EG~J2021+3716} and \object{3EG~J2016+3657} (see
Fig.~\ref{fig:mapa}). These sources could represent the GeV counterparts to the TeV
source \object{MGRO~J2019+37}, which may be a multiple source. Only one of
them, \object{3EG~J2021+3716}, appears in the bright gamma-ray source list  published by the {\it Fermi Gamma-ray Space Telescope}
\citep{abdo09_fermi_bgsl}. Previous observations with {\it AGILE} illustrated its
pulsar nature and inferred an association of this source with
\object{PSR~J2021+3651} \citep{halpern08}.

To explain steady VHE $\gamma$-ray emission, hadronic models have been
developed by several authors \citep[e.g.,][]{aharonian96,
butt03, torres04, bordas09}. The electromagnetic radiation produced by both 
hadronic jets from microquasars and Galactic cosmic rays, and their interaction with the ISM were explored by \cite{bosch05}. The interaction
between the high energy protons, accelerated at the jet termination shock, and
the interstellar hydrogen nuclei produces charged and neutral pions ($\pi^-$,
$\pi^+$ and $\pi^0$); the first set will decay to electrons and positrons and
the second set to photons. The primary radiation, $\pi^0$-decay photons, is
in the $\gamma$-ray band, but the secondary particles can produce significant
fluxes of synchrotron (from radio frequencies to X-rays) and bremsstrahlung
emission (from soft $\gamma$-rays to the TeV range), and in general lower 
efficiency, inverse Compton (IC) emission by interaction with ambient
infrared photons. Detectable fluxes of extended and steady emission should be
produced by this mechanism. Other scenarios involve a jet-driven termination
shock at which relativistic electrons produce synchrotron and TeV IC emission
\citep{aharonian98}. In this context, X-ray observations provide a crucial constraint of the IC emission. 

To understand the nature of the Milagro source in the Cygnus region, we
 performed a multiwavelength campaign comprising a deep radio survey at
610~MHz using the Giant Metrewave Radio Telescope (GMRT) interferometer
covering the 3.5\degr$\times$3.5\degr\ \object{MGRO~J2019+37} field,
near-infrared observations in the $K_{\rm s}$ band using the 3.5~m telescope at
Calar Alto of the central square degree, and archival X-ray data.

This paper is organized as follows. In Sect.~\ref{previous}, we report on
previous radio surveys of the Cygnus region, while in Sect.~\ref{gmrt} we
present the GMRT survey and the results obtained. In Sect.~\ref{nir_survey}, we provide an overview of the near-infrared survey, and in Sect.~\ref{correlation} we
report on the cross-correlations both between our GMRT survey and the near-infrared
survey, and between the GMRT survey and previous X-ray observations. We comment
on particularly interesting sources in Sect.~\ref{individuals} and we discuss  their possible contribution to the TeV emission of \object{MGRO~J2019+37} in
Sect.~\ref{mgro}. We finish with our conclusions in Sect.~\ref{conclusions}.

\section{Previous radio surveys of the Cygnus region} \label{previous}

At radio frequencies, the Cygnus region has been imaged many times, sometimes as
part of Galactic surveys. However, these studies were carried out at poor
angular resolution and/or a relatively high limiting flux density. Some of the
most representative of previous surveys are: the Canadian Galactic Plane Survey
(CGPS) performed with the Synthesis Telescope at the Dominion Radio
Astrophysical Observatory (DRAO) at 408 and 1420~MHz, with angular resolutions
of 5\farcm3 and 1\farcm6, and limiting flux densities of 9 and 1~mJy,
respectively, at declination of +40\degr\ \citep{taylor03}; the Westerbork
Synthesis Radio Telescope (WSRT) 327~MHz survey with an angular resolution of
1\arcmin\ and a limiting flux density of 10~mJy \citep{taylor96}; and the DRAO 408
and 1430~MHz survey with angular resolution of $3.5^\prime\times5.2^\prime $
and $1.0^\prime\times1.5^\prime $, respectively, and limiting flux densities of
150 and 45~mJy, respectively \citep{wendker91}. The most recent survey of this
region is the WSRT 350 and 1400~MHz continuum survey of the Cygnus OB2
association, with angular resolutions of $55^{\prime\prime}$ and
$13^{\prime\prime}$, and limiting flux densities of 10--15 and 2~mJy,
respectively \citep{setia03}. The WSRT survey does not cover the
\object{MGRO~J2019+37} field.

\section{GMRT 610~MHz Radio Survey} \label{gmrt}

\subsection{Observations} \label{gmrt_observations}

The \object{MGRO~J2019+37} region was observed with wide-field deep radio
imaging at 610~MHz (49~cm) using the GMRT, located in Pune (India). We designed
an hexagonal pattern of 19 pointings to cover the region of about
2.5\degr$\times$2.5\degr\ centred on the \object{MGRO~J2019+37} peak of
emission. The observations were carried out in July 2007, but were affected by  a series of power failures in the array and compensatory time was scheduled in August 2007.

The flux density scale was set using the primary amplitude calibrators
\object{3C~286} and \object{3C~48}, which were observed at the beginning and  end of each observing session. On the other hand, phase calibration was
performed by repeated observations of the nearby phase calibrator
\object{J2052+365}. Each pointing was observed for a series of scans 
to achieve a good coverage in the $uv$ plane, the total time spent on
each field being 45~minutes. The total effective time amounts to 20 hours.

Observations were made in two 16-MHz upper and lower sidebands (USB and LSB)
centered on 610~MHz, each divided into 128 spectral channels. The data of each
side-band were separately edited with standard tasks of the Astronomical Image
Processing System ({\tt AIPS}) package. There were no major radio
frequency interference (RFI) problems. However, we did find that narrow band RFI affected a few channels across the band, which were completely flagged. Once
poor antennas, baselines, or channels were removed, the bandpass correction was
used to extend the calibration to all channels. After the bandpass calibration,
the central channels of each sideband were averaged, leading to a data file of
5 compressed channels, of a bandwidth small enough to avoid bandwidth
smearing problems in our images. Standard calibration for continuum data was
performed beyond this point. At the end of the self-calibration deconvolution
iteration scheme, we combined both USB and LSB images of each pointing and
mosaicked the entire region using the {\tt AIPS} task {\tt FLATN}.

We produced different maps of between high and low angular resolution of the GMRT
mosaic. Our highest quality image has an rms of 0.2~mJy~beam$^{-1}$ with a 5\arcsec\
resolution because of the long baselines of the GMRT. A low angular resolution
version was also produced using a restoring beam of 30\arcsec\ to enhance the extended radio sources in the field. This map has an
rms of 0.5~mJy~beam$^{-1}$.

\subsection{Results} \label{gmrt_results}

Figure~\ref{fig:mapa} shows a low angular resolution radio image of the
\object{MGRO~J2019+37} field, together with the position of sources at other
wavelengths. The location of \object{MGRO~J2019+37} is consistent with those of the
EGRET sources \object{3EG~J2016+3657} and \object{3EG~J2021+3716}. The first of
them is positionally coincident with the blazar-like source \object{B2013+370}
(\object{G74.87+1.22}) \citep{mukherjee00,halpern01}, although this blazar is
well outside the inner box of \object{MGRO~J2019+37}. The second is marginally
coincident with the pulsar wind nebula \object{PWN~G75.2+0.1}
\citep{hessels04}. High-energy gamma-ray pulsations originating in the pulsar were detected by {\it AGILE} and {\it Fermi}
\citep{halpern08,abdo09_fermi_bgsl}. There are other known strong and/or
extended radio sources in the field, such as the brightest one inside the
\object{MGRO~J2019+37} center of gravity box, \object{NVSS~J202032+363158}, and
the \ion{H}{ii} region \object{Sh~2-104} (also known as \object{Sh~104}). Other
interesting sources not obvious at first glance become evident when considering
 the whole field in detail. Some of them display a resolved morphology, and in
Sect.~\ref{individuals} we discuss these objects in more detail.

\subsection{Radio catalogue} \label{gmrt_catalogue}

We applied the {\tt SEXtractor} automatic procedure \citep{bertin96} to our
5\arcsec\ resolution mosaic (with pixel size of 1\arcsec) to produce a list of
sources with peak flux density higher than about ten times the local noise
after primary beam correction. Objects with less than 5 connected pixels above
the threshold were not included. The output was visually inspected
and all candidate detections inferred to be false (i.e., deconvolution
artifacts near bright sources) were simply deleted by hand. We used the
local background analysis in SEXtractor to take into account the
uneven background because of beam response effects. Considering the 5\arcsec\ beam
size of the mosaic that we used, and the signal-to-noise ratio that we required for
detection, we estimate that the positions obtained have a typical uncertainty
of 0\farcs5 or smaller.

\begin{figure}
\center
\resizebox{\hsize}{!}{\includegraphics[angle=0]{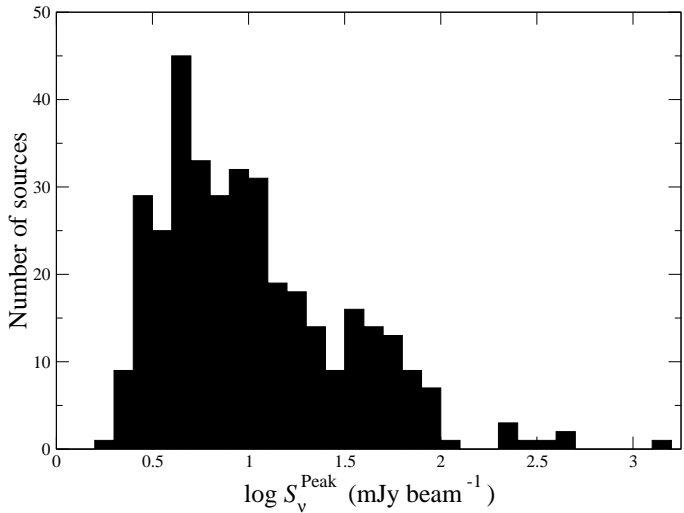}}
\caption{Number of sources versus~$\mbox{log}\ S_\nu^{\rm Peak}$ for the 362
sources detected in the GMRT 610~MHz radio survey.}
\label{fig:histo}
\end{figure}

The resulting list, considered to be very reliable although not complete at 
the lowest flux density levels, contains 362 radio sources. Among them, 203 are
fainter than 10~mJy and the majority were previously undetected at radio
wavelengths. We present the catalogue in Table~\ref{table:list} of the online
material accompanying this paper. The first and second columns provide the
catalogue number and the source name. The third and fourth columns give the
J2000.0 position in right ascension order. The fifth and sixth columns provide
the peak flux density and the local noise, respectively. The seventh and eighth
columns list the integrated flux density and its error. Uncertainties quoted
for the peak and integrated flux densities are based on the formal errors of
the fit and allow the reliability of the detection to be judged. However,
they do not include the error due to primary beam correction as a
function of angular distance to the phase centre because of unknown antenna
offsets, which is estimated to be around 10\% of the flux density values
\citep[see for instance][]{paredes08}. In Fig.~\ref{fig:histo}, we show the
source distribution histogram as a function of $\mbox{log}\ S_\nu^{\rm Peak}$. 

\section{Near-infrared survey} \label{nir_survey}

We also carried out a near-infrared (NIR) survey of the central square
degree of the region using the OMEGA2000 wide field camera
($15\arcmin\times15\arcmin$) on the 3.5~m telescope at Centro Astron\'omico
Hispano Alem\'an (CAHA) in Calar Alto (Spain) on 25 September 2007. This
instrument consists of a Rockwell HAWAII2 HgCdTe detector with $2048\times2048$
pixels sensitive from 0.8 to 2.5 $\mu$m. The observations were performed in the
$K_{\rm s}$-band (2.15 $\mu$m) to minimize the interstellar
absorption. Individual frames were sky-subtracted, flat-field corrected, and
then combined into a final mosaic using the {\tt AIPS} task {\tt FLATN}. The
ensamble of $4 \times 4$ pointings covers almost completely the center of
gravity and uncertainty region of the TeV emission from the source
\object{MGRO~J2019+37}. The average limiting magnitude across the mosaic is
$K_{\rm s}\simeq 17$~mag, and the total field of view is
$0.9\degr\times0.9\degr$. Astrometric solutions for the final frames were
determined within $\pm$0\farcs1 by identifying about twenty reference stars in
each pointing, for which positions were retrieved from the 2MASS catalogue
\citep{skrutskie06}. A catalogue of $\sim$315\,000 near-infrared sources was
produced using the SEXtractor package.

\section{Radio, near-infrared, and X-ray cross-correlation catalogue} \label{correlation}

We performed a cross-correlation between the radio and near-infrared source
catalogues. Considering the 0\farcs5 uncertainty in the radio positions and the
0\farcs1 uncertainty in the NIR ones, we used a conservative maximum
offset of 0\farcs6 for associations (neglecting systematics between both
catalogues). There are 42 of the 362 detected radio sources inside the area
imaged in the near infrared. A total of 6 of these 42 sources have a
near-infrared counterpart candidate within 0\farcs6 of their radio position.
Their magnitudes are listed in the ninth column of Table~\ref{table:list} of
the online material. The chance coincidence probability of finding a NIR source
closer than 0\farcs6 to a given radio source is estimated to be the number of NIR
sources multiplied by the area of the uncertainty in positions occupied by the
42 radio sources, divided by the total area of the region:
$(315\,000\times42~\pi~0\farcs6^2) / (3240\arcsec\times3240\arcsec)=1.4$.
Therefore, of the six radio sources with NIR counterpart, we expect that one of
them is a random coincidence. 

We also obtained source lists of all X-ray observations of the region
performed by {\it Chandra} and {\it XMM-Newton}, computed by the
\verb+celldetect+ and \verb+edetect_chain+ tasks from CIAO~4.0 and SAS~8.0, 
respectively. A total of 41 of the 362 radio sources are located in fields
observed in X-rays, which cover an area of $314$~arcmin$^2$
($1\,130\,973$~arcsec$^2$) and contain 519 X-ray sources. We found that 5
of the 41 radio sources have an X-ray counterpart candidate within 5\arcsec\
(the typical uncertainty for {\it XMM-Newton}). Their X-ray fluxes are listed
in the tenth column of Table~\ref{table:list} of the online material. The
chance coincidence probability of finding a radio source closer than 5\arcsec\ to a
given X-ray source is estimated to be the number of radio sources multiplied by
the area of the uncertainty in positions occupied by the 5 X-ray sources,
divided by the total area of the region: $(41\times519~\pi~5\arcsec^2) /
(1\,130\,973$~arcsec$^2)=1.4$. Therefore, of the five X-ray sources with radio
counterpart we also expect that one of them is a random coincidence. 

A single triple radio/near-infrared/X-ray coincidence has been found (source
number 115 in Table~\ref{table:list} of the online material).

\section{Individual sources in the \object{MGRO~J2019+37} field} \label{individuals}

The most interesting radio sources that appear in the uncertainty region of the
TeV emission (red box in Fig.~\ref{fig:mapa}) are described below.

\subsection{\object{PSR~J2021+3651} / \object{PWN~G75.2+0.1}} \label{pwn}

The radio pulsar \object{PSR~J2021+3651} has a rotation period $P=0.104$~s, a
characteristic age of $P/2\dot{P}=17$~kyr, and a spin-down luminosity
$\dot{E}=3.4\times10^{36}$~erg~s$^{-1}$. It is coincident with the unidentified
source \object{GeV~2020+3658} \citep{roberts02}, which overlaps with the EGRET
source \object{3EG~J2021+3716} (see Fig.~\ref{fig:mapa}). {\it Chandra}
observations of this pulsar detected a $\sim 20\arcsec\times 10\arcsec$ pulsar
wind nebula named \object{PWN~G75.2+0.1} \citep{hessels04}. {\it
Chandra} observations of the pulsar and its PWN detected rings and jets around
\object{PSR~J2021+3651}, and inferred a distance to the pulsar of 3--4~kpc
\citep{vanetten08}, in contrast to the 12~kpc implied by the pulsar
dispersion measure \citep{roberts02}. {\it XMM-Newton} observations show
emission extending to a distance of $\sim$10--15 arcminutes, whereas radio observations with
the VLA at 1.4~GHz show a radio nebula coincident with the X-ray extension
\citep{roberts08}.

{\it AGILE} detected the source \object{AGL~J2020.5+3653} at energies above
100~MeV range, which shows pulsations and was associated with the pulsar
\object{PSR~J2021+3651} \citep{halpern08}. The photon spectrum of the source
can be fitted by a power-law of photon index $\Gamma=1.86\pm0.18$ in the
range 100--1000~MeV, while a turndown is seen above 1.5~GeV. This source
appears as \object{1AGL~J2021+3652} in the first {\it AGILE} catalog of high
confidence gamma-ray sources \citep{pittori09}. {\it Fermi} also detected
the source \object{0FGL~2020.8+3649} in positional coincidence with the pulsar
\citep{abdo09_fermi_bgsl}. 

We found neither a (low-frequency) radio nor a near-infrared source at the
position of \object{PSR~J2021+3651}. The nearest near-infrared source is at a
distance of 3\farcs9 and has a $K_{\rm s}$ magnitude of 17.3. In the
radio, from our 610~MHz GMRT data we can establish an upper limit to
any possible point-like counterpart of 1.0~mJy by multiplying the background
emission level by a factor of 5. The radio flux density of the extended emission found with the
VLA at 1.4~GHz amounts to $\sim$700~mJy in an area of about 100~arcmin$^2$,
which for a uniform distribution yields 7~mJy~arcmin$^{-2}$. On the other hand,
the rms of our low-resolution radio map at 610~MHz shown in
Fig.~\ref{fig:mapa}, with a beam size of 30\arcsec, is 0.5~mJy~beam$^{-1}$.
This provides a conservative 5-$\sigma$ upper limit of either 2.5~mJy~beam$^{-1}$ or
9~mJy~arcmin$^{-2}$. This upper limit implies that if the radio emission is
uniformly distributed, its spectral index must be above $-$0.3. This value is
compatible with the radio emission being produced by the synchrotron mechanism,
as expected in this nebula.

\subsection{Jet-like radio sources} \label{jet}

\begin{figure}[h!]
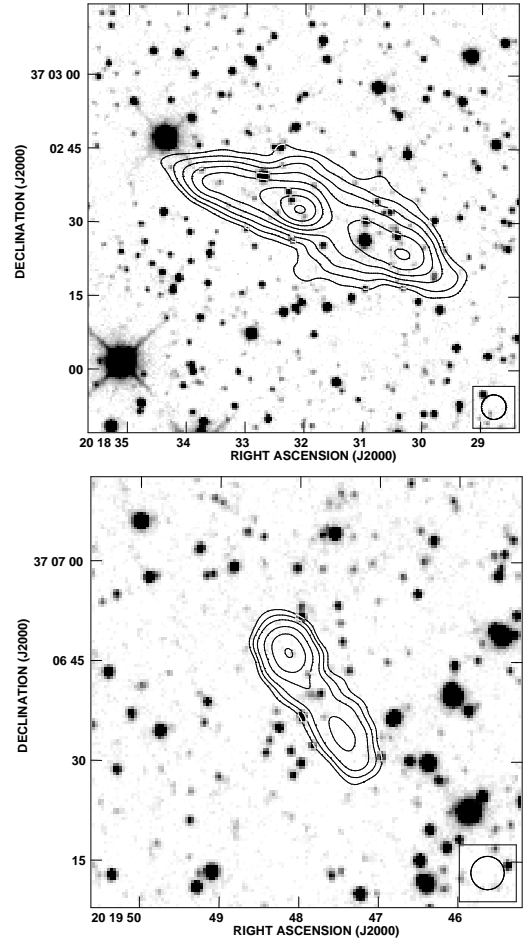

\center
\resizebox{0.75\hsize}{!}{\includegraphics[angle=0]{f3a.eps}}
{\vspace{1mm}}
\resizebox{0.75\hsize}{!}{\includegraphics[angle=0]{f3b.eps}}
\caption{Radio and near-infrared image composition of the jet-like radio
sources A (top) and B (bottom). The GMRT radio contours are superimposed on the
$K_{\rm s}$-band 3.5~m CAHA telescope images. {\it Top}: Source A. Contours 
correspond to 5, 9, 15, 23, 45, 60, and 80 times 0.16~mJy~beam$^{-1}$, the rms
noise. The integrated flux density of the source is 164.7$\pm$0.3~mJy.
{\it Bottom}: Source B. Contours correspond to 5, 8, 12, 20, 30, and 42 times
0.16~mJy~beam$^{-1}$, which is the rms noise. The integrated flux density of the
source is 28$\pm$0.1~mJy. The synthesized radio beams of 5\arcsec\ are plotted
in the lower-right corners of both images.}
\label{fig:jets}
\end{figure}

We discovered two jet-like radio sources located well inside the
uncertainty region of \object{MGRO~J2019+37}. Their J2000.0 positions are
$\alpha$=$20^{\rm h}18^{\rm m}32^{\rm s}$, $\delta$=+$37^{\circ}02\farcm5$
(source A) and $\alpha$=$20^{\rm h}19^{\rm m}48^{\rm s}$,
$\delta$=+$37^{\circ}06\farcm7$ (source B). In Fig.~\ref{fig:jets}, we show  a
GMRT high resolution image of each of them superimposed on the near-infrared
image. Both sources appear to be unresolved in the NVSS 1.4~GHz catalogue
\citep{condon98}. Based on our GMRT survey, the NVSS survey, and the Westerbork Northern
Sky Survey (WENSS; \citealt{rengelink97}) data at 327~MHz, we estimate a
spectral index of $-$1.16$\pm$0.02 for source A, clearly indicating a
non-thermal nature. It is interesting to note that the source is not detected
in the VLA Low-Frequency Sky Survey (VLSS; \citealt{cohen07}) at 74~MHz, with a
3-$\sigma$ upper limit of 1.2~Jy. With the spectral index above, we would expect
a flux density of 1.9~Jy, clearly indicative of a turnover at
lower radio frequencies, which could be produced by intrinsic self-absorption
or by Galactic foreground free-free absorption. The GMRT and NVSS data for
source B provide a spectral index of $-$0.7$\pm$0.6, compatible with the
non-detection in WENSS, and suggesting a non-thermal nature for this radio
source.

Source A (Fig.~\ref{fig:jets}-top) shows a double-sided morphology,
sources \#141 and \#142 in Table~\ref{table:list} of the online material, with
a slight bending towards the south-east. This structure resembles ones typically 
seen in radio galaxies with a non-negligible pressure from the intergalactic
medium. Unfortunately, there is no clear extended NIR counterpart in the axis
joining the radio lobes that could be identified with the parent galaxy, and no
firm conclusion can be obtained from the present data.

\begin{figure*}
\center
\resizebox{1.00\hsize}{!}{\includegraphics[angle=0]{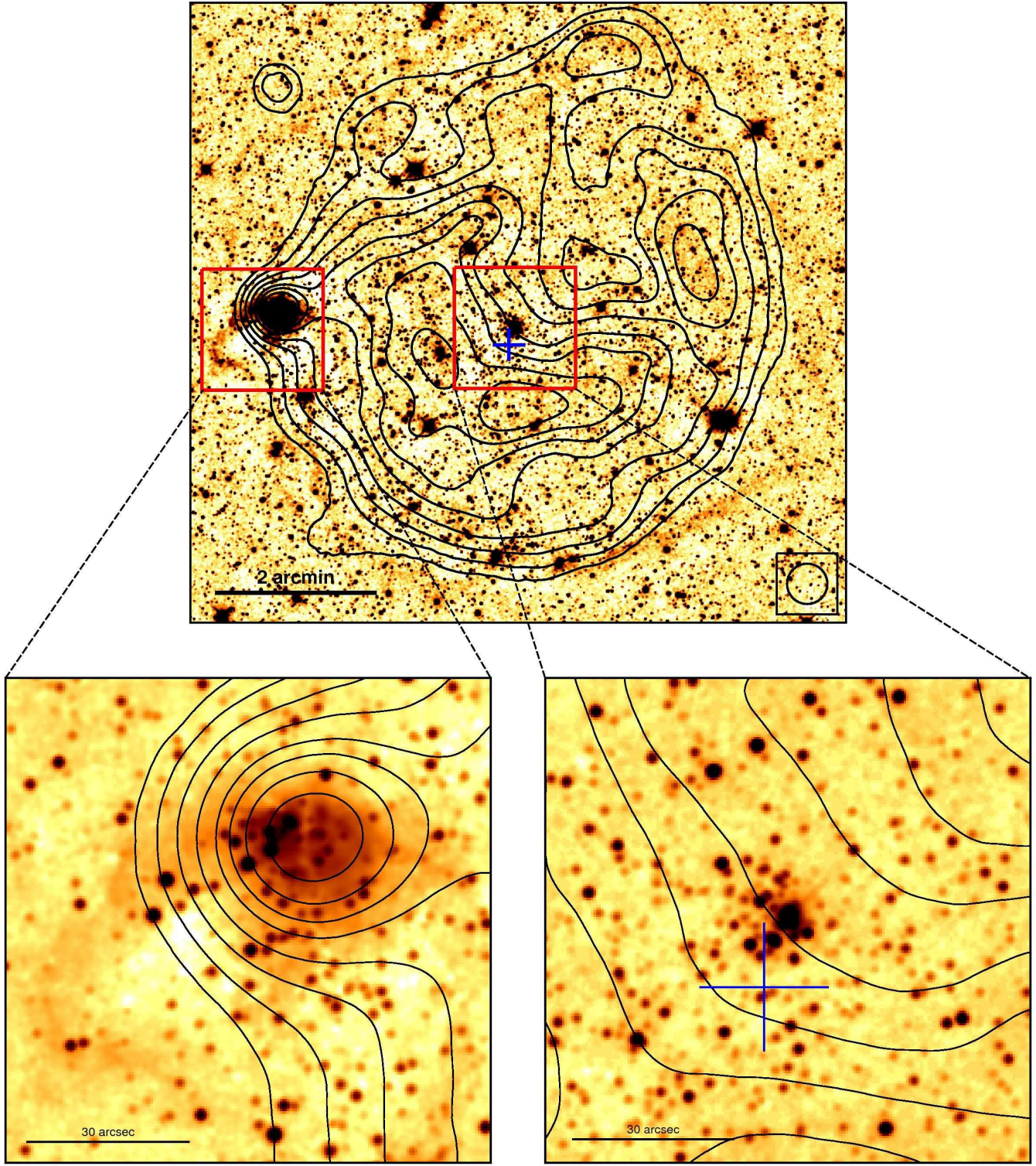}}
\caption{{\it Top}: Composite radio and near-infrared image centred on the
\object{Sh~2-104} region. The contours correspond to 10, 20, 35, 55, 80,
100, 125, 155 times the rms noise of 0.3~mJy~beam$^{-1}$ of our GMRT 610~MHz
(49~cm wavelength) image. We overlay our $K_{\rm s}$-band near-infrared image
of the same field obtained using the 3.5~m telescope at Calar Alto. The blue
cross marks the position and 1-$\sigma$ uncertainty of the {\it ROSAT} source
\object{2RXP~J201742.3+364513}. {\it Bottom-left}: Young massive stellar
cluster deeply embedded in a UCHII region found on the eastern rim of
\object{Sh~2-104}. {\it Bottom-right}: New cluster candidate in the center of
\object{Sh~2-104} previously assumed to be a single star (identified later as
\object{2MASS~J20174184+3645264}). The scale of each image is indicated by the
horizontal bar. The colour scale of the bottom images was changed to display the individual stars within each cluster more clearly.}
\label{fig:sh104}
\end{figure*}

Source B (Fig.~\ref{fig:jets}-bottom) shows a morphological and spectral
similarity to the radio lobes of the `great annihilator'
\object{1E~1740.7$-$2942}, a microquasar at the Galactic center
\citep{mirabel92}. The two lobes correspond to sources \#193 and \#194 in
Table~\ref{table:list} of the online material. We did not detect a radio
core in this source but, as for the one present in \object{1E~1740.7$-$2942}, it
could have a flat spectrum and the flux density at such a low frequency is
expected to be very low compared to that of the radio lobes. As can be seen in the
figure, there are two near-infrared objects close to the central position of
the source. Their J2000.0 coordinates and magnitudes are: $\alpha$=$20^{\rm
h}19^{\rm m}47\fs74$, $\delta$=+$37^{\circ}06^{\prime}40\farcs2$, $K_{\rm
s}$=16.5, and $\alpha$=$20^{\rm h}19^{\rm m}47\fs86$, 
$\delta$=+$37^{\circ}06^{\prime}39\farcs9$, $K_{\rm s}$=17.4. Their proximity
significantly biases the photometry. The bright source is point-like and offset
from the axis traced by the radio lobes. The faint source is aligned with the axis and fuzzy, implying that the origin of the double radio source is most likely a radio galaxy.

Previous {\it ROSAT} pointed observations (Obs. Id. 500248P conducted on 24
October 1993) did not detect any of these two radio-jet sources, placing a
3-$\sigma$ upper limit of $7\times10^{-14}$ erg~cm$^2$~s$^{-1}$ on their
persistent flux in the energy range 0.1--2.4~keV. With the present data, we
cannot elucidate whether the sources are Galactic or extragalactic, although
there are hints of their extragalactic nature.

\subsection{\ion{H}{ii} region \object{Sh~2-104}} \label{hii}

\object{Sh~2-104}, also known as \object{Sh~104}, is an optically visible
\ion{H}{ii} region of $7^\prime$ diameter at a distance of 4.0$\pm$0.5~kpc
\citep{deharveng03}. There is a central O6\,V star suspected of being 
responsible for ionizing the region \citep{lahulla85}. The appearance of
\object{Sh~2-104} in the optical and in the radio bands is very similar,
although the radio images show the presence of an ultra compact \ion{H}{ii}
(UCHII) region at the eastern border, which is not visible in the optical image
\citep{deharveng03}. The interaction between the expanding \ion{H}{ii} region
\object{Sh~2-104} and the UCHII region may be responsible for triggered star
formation in the latter, resulting in a deeply embedded young cluster. This
region has also been detected as a high luminosity ($3\times10^4$ L$_{\odot}$)
{\it IRAS} source.

Our GMRT observations (see Fig.~\ref{fig:sh104}) detect a structure similar to
that found at 1.46~GHz with the VLA \citep{fich93} and at 1.4~GHz within the
NVSS radio continuum survey \citep{condon98}.


We also observed \object{Sh~2-104} in the near-infrared $K_{\rm s}$-band.
The images obtained are deeper than those from 2MASS. Figure~\ref{fig:sh104}
shows our near-infrared images of the field of \object{Sh~2-104} with the radio
emission contours superimposed. In the eastern region of the ring (to the left
side), the near-infrared image shows the well known cluster associated with the
UCHII region, which must contain at least one massive OB star
\citep{deharveng03}. In the central part of the image, the single O6\,V star of
\cite{lahulla85}, which corresponds to \object{2MASS~J20174184+3645264}
\citep{skrutskie06}, now appears to be resolved as several point-like objects,
indicative of the presence of a cluster. Therefore, apart from this ionizing early-type star, the new cluster candidate could also contribute to the formation of
the \ion{H}{ii} region (e.g., additional early type stars, wind shocks).
Furthermore, an elongated arc along the south of \object{Sh~2-104} as well as
to the east of the UCHII region can be discerned in the NIR images. These
features may be related to the interaction between the expanding \ion{H}{ii}
region and the interstellar medium.

Despite its deep coverage at other wavelengths, \object{Sh~2-104} was 
poorly explored in the X-ray domain. Previous X-ray observations of this region
by {\it ROSAT} detected a source (\object{2RXP~J201742.3+364513};
\citealt{rosat00}) located at $\alpha$=$20^{\rm h}17^{\rm m}42\fs3$,
$\delta$=+$36^{\circ}45^{\prime}13\arcsec$ with a positional error of
$\sim$12\arcsec, overlapping with the central star
\object{2MASS~J20174184+3645264} and the cluster candidate (see bottom-right of
Fig.~\ref{fig:sh104}). The count rate of
$(4.1\pm0.5)\times10^{-3}$~count~s$^{-1}$ in the energy range 0.1--2.0~keV,
provides a flux of $(5.8\pm0.7)\times10^{-14}$~erg~cm$^{-2}$~s$^{-1}$ based on  the
assumption of a thermal spectrum with a temperature of 1.5~keV (a typical value
for colliding wind regions). On the other hand, OB stars are known to be X-ray
sources, presumably because of shocks in their stellar winds (see
\citealt{guedel04} for a review). According to the complete study by
\cite{berghoefer97} of more than 200 isolated OB stars detected in {\it ROSAT} data,
for an O6\,V star, with bolometric luminosity of $8\times10^{38}$~erg~s$^{-1}$
\citep{martins05}, the corresponding X-ray luminosity is
$1.2\times10^{32}$~erg~s$^{-1}$. Considering a distance of 4.0~kpc to both 
\object{Sh~2-104} and its ionizing central star
\object{2MASS~J20174184+3645264}, the expected X-ray flux is
$6.3\times10^{-14}$~erg~cm$^{-2}$~s$^{-1}$, fully compatible with the detected
X-ray flux from \object{2RXP~J201742.3+364513}.

\subsection{\object{NVSS~J202032+363158}} \label{nvss}

\begin{figure}
\center
\resizebox{\hsize}{!}{\includegraphics[angle=0]{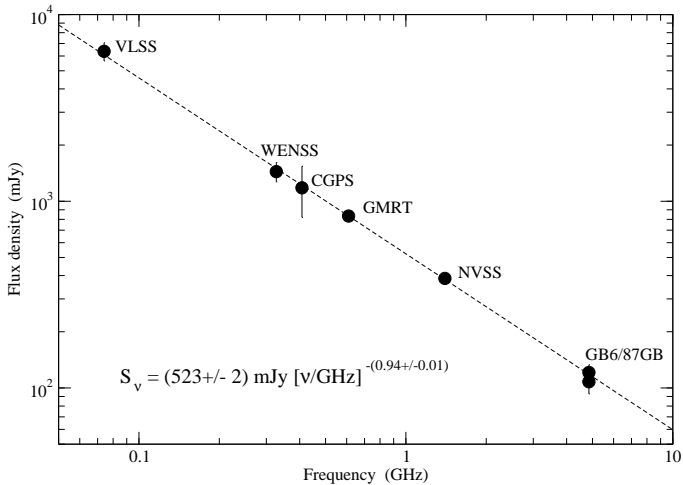}}
\caption{Radio spectrum of \object{NVSS~J202032+363158} based on the flux
densities compiled in Table~\ref{table:flux}. The straight line is a simple
power-law fit.}
\label{fig:spec}
\end{figure}

The source \object{NVSS~J202032+363158} is the brightest compact radio source
within the error box of the TeV peak emission of \object{MGRO~J2019+37} in our
GMRT observations. Its flux density at 610~MHz is 833~mJy and has not been
resolved. This source appears in the VLSS at 74~MHz, in the WENSS at 327~MHz,
in the CGPS at 408~MHz \citep{taylor96}, in the Effelsberg survey of the
Cygnus~X region at 1420~MHz \citep{wendker91}, and in the Green Bank 4.85~GHz
northern sky surveys 87GB \citep{gregory91} and GB6 \citep{gregory96}. In Table~\ref{table:flux}, we
summarize  the detected flux densities within these
surveys, and we plot the corresponding spectrum in Fig.~\ref{fig:spec}.
Assuming a stable flux density, the radio spectrum of this source can be
described by $S_{\nu} = (523\pm 2)~{\rm mJy}~[\nu/{\rm GHz}]^{-0.94\pm 0.01}$, 
and it is, therefore, clearly a non-thermal emitter. Despite very low
frequencies being sampled, no evidence of the turnover frequency below 1~GHz is
obvious from this simple power-law fit.

\begin{table}
\begin{center}
\caption[]{Non-simultaneous flux density measurements of the source
\object{NVSS~J202032+363158} obtained from different surveys.}
\begin{tabular}{lcc}
\hline
\hline 
Survey & Frequency & Flux density \\
       & (MHz)     & (mJy) \\
\hline
VLSS   & \phantom{11}74 &           6354$\pm$708 \\
WENSS  & \phantom{1}327 &           1442$\pm$216 \\
CGPS   & \phantom{1}408 &           1180$\pm$360 \\
GMRT   & \phantom{1}610 & \phantom{1}833$\pm$\phantom{1}56 \\
NVSS   &           1400 & \phantom{1}386$\pm$\phantom{1}12 \\
GB6    &           4850 & \phantom{1}121$\pm$\phantom{1}12 \\
87GB   &           4850 & \phantom{1}108$\pm$\phantom{1}15 \\
\hline
\label{table:flux}
\end{tabular}
\end{center}
\end{table}

\begin{figure*}[ht!]
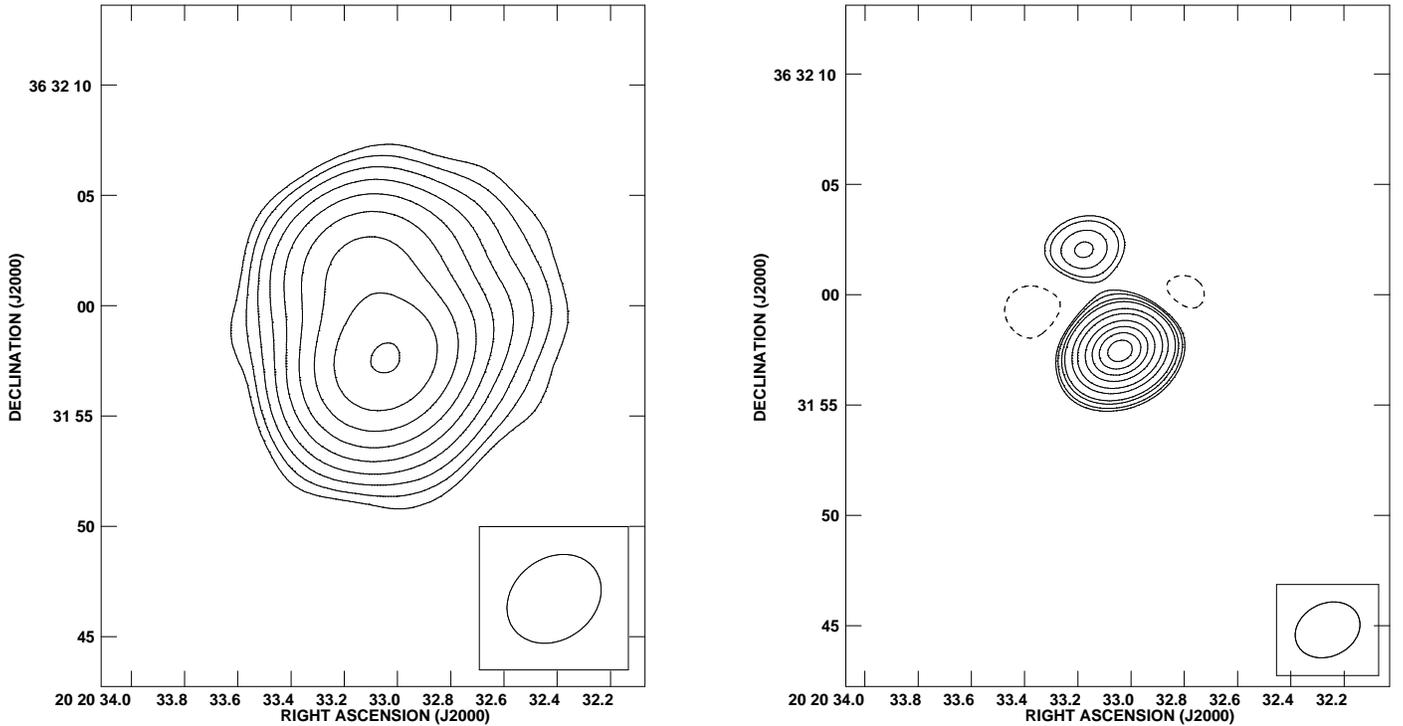

\center
\resizebox{1.0\hsize}{!}{\includegraphics[angle=0]{f6a.eps}\hspace{3cm}\includegraphics[angle=0]{f6b.eps}}
\caption{\emph{(Left)}: Image of the source \object{NVSS~J202032+363158} at
21~cm obtained using uniform weights on B-configuration VLA data. The source is
resolved, displaying a one-sided radio jet extending a few arc-sec towards the
north, with a core component of $\sim$250~mJy and a secondary component of
about 70~mJy. The rms of the image is 0.22~mJy~beam$^{-1}$. Contours correspond
to 4, 8, 16, 32, 64, 128, 256, 512, and 1024 times the rms noise.
\emph{(Right)}: Image from the same data performed using an $uv$-range of
30--50~k$\lambda$, which clearly shows the core and the component discussed in
the text, with peak flux densities of 170 and 20~mJy, respectively. The rms of
the image is 2.5~mJy~beam$^{-1}$. Contours correspond to $-$3, 3, 4, 6, 8, 14,
20, 30, 40, 50, and 60 times the rms noise. The synthesized radio beams are
plotted in the lower-right corners of both images.}
\label{fig:nvss}
\end{figure*}

By inspecting of the NRAO archives, we found a previous VLA snapshot (6 min on
source) of this radio source at the 20~cm wavelength in B configuration 
(providing a nominal synthesized beam of 4\arcsec) observed on 25 March 1989.
This observation was calibrated using standard {\tt AIPS} tasks, 
including phase self-calibration. A uniformly weighted image is shown in
Fig.~\ref{fig:nvss}-left. As can be seen, this radio source is resolved,
displaying a one-sided radio jet extending a few arcsec towards the north,
with a core component of $\sim$250~mJy and a secondary component of about
70~mJy. To enhance the compact structure of the source, we obtained an
image for the longest baselines of the same VLA run, using a $uv$-range of
30--50~k$\lambda$. The image, shown in Fig.~\ref{fig:nvss}-right, clearly shows
a compact core and a secondary component, with peak flux densities of 170 and
20~mJy, respectively, resembling the large-scale jet of a microquasar.

To explore the source at higher angular resolutions, we observed the core of
\object{NVSS~J202032+363158} at 1.6~GHz (18~cm wavelength) with the European
VLBI Network in eVLBI mode (eEVN). This is a technique in which the signals from
distant radio telescopes are directly streamed into the central data processor
for real-time correlation, instead of being recorded on disk or tape. The
observation took place on 3 March 2007 from 5:00 to 13:00~UT (centered on
MJD~54163.375), and was performed using 6 antennas: Cm, Mc, Jb-2, On-85, Tr,
and Wb. Scans on \object{NVSS~J202032+363158} were interleaved with scans on
the compact phase calibrator J2015+3710, with a 6-min cycle time (66~s on
the calibrator and 246~s on the source). The data were recorded using dual
polarization and 2-bit sampling, at 256~Mbps. A total bandwidth of 32~MHz per
polarization was provided by 4 sub-bands. The e-VLBI data were processed at the
Joint Institute for VLBI in Europe (JIVE) correlator in real time, using an
integration time of 2 sec. The target source was correlated with the position
obtained from the VLA-B (30--50~k$\lambda$) data: $\alpha_{\rm J2000.0}=20^{\rm
h} 20^{\rm m} 33\fs0401$ and $\delta_{\rm J2000.0}=+36\degr 31\arcmin
57\farcs480$, for a total maximum uncertainty of 100~mas. During observations, we experienced 
synchronization problems  and the correlation had to be
restarted several times. A few antennas were dropped out of the correlation
jobs during the gaps used for measuring the system temperatures. Due to these
disconnections, part of the data, which is not recorded onto disks for these
experiments, was lost during the correlation, and the true on-source time is
estimated to be around 3~hours.

We performed the post-correlation data reduction using the {\tt AIPS}
software package and {\tt Difmap}. We applied ionospheric corrections to the
visibility data, and the system temperatures were used to obtain the a priori
visibility amplitude calibration. All stations produced fringes with the 1-Jy
phase calibrator, situated at 1\fdg2 from the target, and we therefore  transferred
the solutions for the phases to the target source. We improved the amplitude
calibration using correction factors for each antenna obtained from the
self-calibration of J2015+3710. Self-calibration of the
\object{NVSS~J202032+363158} data was impossible because of the lack of bright
sources in the primary beam of the antennas. The phased-referenced
natural-weighted image that we obtained had a synthesized beam of 22.7$\times$19.8~mas at a 
position angle of 30\fdg3, and an rms noise of 0.20~mJy~beam$^{-1}$. No
significant detections were found within a distance of 5\arcsec\ from the
correlated phase center.

There is no near-infrared counterpart candidate to
\object{NVSS~J202032+363158}. The nearest sources are both at 4\farcs1, with
magnitudes of 14.2 and 17.3 in the $K_{\rm s}$-band.

\section{Could any of the selected individual sources power the TeV emission
from \object{MGRO~J2019+37}?} \label{mgro}

\object{MGRO~J2019+37} covers a sky region of approximately 
$1\degr\times 1\degr$. The extended emission could be produced by either a
single powerful accelerator, or by the superposition of several point-like
sources. Although we focus on the individual sources presented in
Sect.~\ref{individuals}, we cannot exclude some of the 
additional radio sources listed in Table~\ref{table:list} of the online
material being responsible for, or contributing to, the Milagro source.

If \object{MGRO~J2019+37} is a single extended source, and not a
combination of different sources, the origin of the $>$12~TeV emission is
likely to be hadronic. The time required to fill a region of a size of $\sim$1\degr\
(or (1--5)$\times10^{20}$~cm at 2--10~kpc distance) with electrons of
$\sim$100~TeV by means of diffusion is
\begin{equation}
t_{\rm diff}=1.5\times10^{12}~R_{20}^2~B_{-6}~~{\rm s},
\end{equation}
where $R_{20}=R/10^{20}$~cm is the source size, and $B_{-6}=B/10^{-6}$~G is the
ISM turbulent magnetic field. For realistic ISM densities of $n_{\rm
ISM}<10^4$~cm$^{-3}$, and reasonable magnetic/mm-far IR field energy densities,
i.e., 
$>$1~eV~cm$^{-3}$, the electron cooling timescale is dominated by synchrotron
and IC losses and found to be $t_{\rm cool}<10^{11}$~s. Therefore, given that
$t_{\rm diff}\gg t_{\rm cool}$, electrons injected from a single accelerator
cannot fill the entire multi-TeV source. Otherwise, protons cool mainly by means of 
collisions with the ISM nuclei ($pp$)
\begin{equation}
t_{\rm cool}\sim 10^{15}/n_{\rm ISM}~~{\rm s}.
\end{equation}
For $n_{\rm ISM}\sim$1\,000--200~cm$^{-3}$ (2--10~kpc) in the Milagro region,
proton injection luminosities of $\sim 10^{37}$~erg~s$^{-1}$ should be enough
to explain the observed luminosities \citep{abdo07_cygnus} assuming that
$\sim$0.1--1\% of the proton power is in $>$12~TeV photons. As a result of $pp$
interactions, secondary electron-positron pairs and neutrinos are also produced  with luminosities and energies similar to those of gamma-rays
\citep[e.g., ][]{kelner06}. These secondary pairs should radiate via synchrotron,
relativistic bremsstrahlung, and IC. Extended radio and X-ray emission was
detected within the Milagro region \citep{roberts08,hessels04,vanetten08}.
However, the smaller extent of these diffuse sources compared to the size of
the TeV emission makes any possible association difficult. Nevertheless, for
typical ISM densities and magnetic and radiation fields, most of the emission
from the secondary pairs could be produced at relatively low gamma-ray fluxes, rendering them undetectable.

Once the most probable emission scenario is decided, we will be able to see  whether the different objects 
proposed in Sect.~\ref{individuals} could act as the accelerator. 

The spin-down luminosity of \object{PSR~J2021+3651} is marginally in agreement
with the energetic requirements stated above. Nevertheless, for this object to
act as the accelerator, most of this luminosity should be in the form of protons
\citep[as in, e.g.,][]{horns07b}. In addition, the accelerated protons should
escape the $\sim 10'$ nebula in a time shorter than or equal to the age of the
pulsar, $\approx 17$~kyr, which may not be possible if the turbulent magnetic
field in the nebula reaches value of several 10~$\mu$G or higher. On the other
hand, the turndown in the {\it AGILE} GeV spectrum questions the association of
\object{AGL~J2020.5+3653} as the only counterpart to \object{MGRO~J2019+37}. Extrapolation of the last two data points in the spectrum shown
in \cite{halpern08} provide a flux at 20~TeV a factor of 3500 below the
reported \object{MGRO~J2019+37} flux \citep{abdo07_survey}. Even ignoring the
turndown and fitting the entire spectrum with a single power-law, there is still
a one order of magnitude difference. Therefore, if no additional components are
present in the GeV-TeV spectrum of
\object{PSR~J2021+3651}/\object{PWN~G75.2+0.1}, this source alone can hardly
explain the multi-TeV emission from \object{MGRO~J2019+37}.

The massive star and the 
star-forming region (MSR; SFR) associated with the \ion{H}{ii} region
\object{Sh~2-104} could be responsible for the extended Milagro source if  they
were capable of injecting $\sim$10$^{37}$~erg~s$^{-1}$ in the form of relativistic
protons into their surroundings. Assuming an efficiency of a 10\% for the kinetic
energy converted to non-thermal proton energy in the shocks present inside the
\ion{H}{ii}/SFR region, about 100 massive (proto)stars
producing jets or winds with velocities of $\sim 10^8$~cm~s$^{-1}$ 
and mass-loss rates of $\sim 10^{-6}~{\rm M}_{\odot}~{\rm
yr}^{-1}$ would be required to reach the needed proton luminosities. It seems
unlikely that \object{Sh~2-104} can harbor such a high number of massive 
(proto)stars.
However, \object{Sh~2-104} may be part of a larger MSR or 
SFR that have not yet 
been detected, and in that case, the larger whole MSR or SFR could represent  the emitter
of the whole Milagro source through wind collisions or jet/medium interactions 
\citep[see, e.g.,][]{torres04,romero08}, respectively. In
this scenario, thermal free-free radio emission from the whole SFR would be
expected. The non-detection of this SFR in our GMRT observations could be
explained by free-free absorption in the ionized regions and the surrounding material of 
the MSR/SFR, 
although the development of the particular details of this scenario are beyond the scope of this work.
Observations searching for maser emission with instruments such as Apex or
Nanten could help us to detect this hypothetical star-forming region.

We note that the accelerator itself might be outside \object{MGRO~J2019+37},
as in the case of the stellar cluster \object{Berkeley~87} mentioned in
\cite{abdo07_cygnus}. This cluster could accelerate the protons that would
then escape from it diffusing towards, and ultimately interacting with, a denser
region located near the Milagro source best-fit model position.

We found three non-thermal radio sources with jet-like structures in
the field of \object{MGRO~J2019+37}: sources A and B, and source
\object{NVSS~J202032+363158}. Although some arguments support the
extragalactic nature of sources A and B, we cannot exclude the possibility that they are Galactic in nature. VLBI observations of \object{NVSS~J202032+363158} provide an upper limit
of 1~mJy~beam$^{-1}$ to the flux for a beam size of $\sim$20~mas. Therefore, this source
did not exhibit a compact core during our observations. It was either completely
resolved or is a variable radio source, since no radio emission is expected in the
high/soft state of microquasars. In any case, these three sources could be
hadronic microquasars whose jets would interact with the ISM accelerating
protons \cite[e.g.,][]{heinz02,bordas09}. The accelerated protons may escape
from the accelerating region colliding with the surrounding regions of the ISM,
rendering very high-energy emission \citep[e.g.,][]{bosch05}. From the
energetic point of view, although these sources could explain the Milagro source, the
lack of clear X-ray counterparts needs to be explained, if accretion is taking
place in these objects. It might be the case that accretion is inefficient in
producing X-rays (as could be the case in \object{LS~5039}; e.g., 
\citealt{bosch07}). Finally, a microquasar located outside the Milagro region
could be powering the multi-TeV radiation.

The constraint on a hadronic origin for the Milagro emission does not apply if
the source consists of different accelerators/emitters. In this case,
several leptonic emitters, which may or may not coincide with (some of) the
sources discussed here, could be behind \object{MGRO~J2019+37}. 

From this analysis, we conclude that several objects should be considered when 
trying to understand the origin of \object{MGRO~J2019+37}, although the nature
of the accelerator/emitter remains uncertain. Insights into this
question could be provided by further multiwavelength studies, by future
imaging atmospheric Cherenkov telescopes (MAGIC-II, H.E.S.S.-II), and by $\it
{Fermi}$, which should be able to constrain the source position and
morphology more tightly, and explore in detail its physics by obtaining  spectral
information across a broad wavelength range. Finally, neutrino detections with future neutrino instruments
could provide additional evidence to support the hadronic scenario.

\section{Conclusions} \label{conclusions}

We have carried out a deep radio survey of about 6 square degrees region in the
direction of \object{MGRO~J2019+37}, and a near-infrared survey of the central
square degree. This has provided a catalogue of 362 radio sources and a
catalogue of 315\,000 NIR sources. The radio and NIR data presented here detect
a large number of previously unknown sources and shed additional light on known
objects. We have found that if a single accelerator is powering
\object{MGRO~J2019+37}, the most likely origin of the multi-TeV emission
is hadronic in nature. We have shown that the extrapolation of the spectrum of
the pulsar \object{AGL~J2020.5+3653} does not explain the detected flux from
\object{MGRO~J2019+37}. This indicates either that there is an additional
component in the GeV-TeV spectrum of the pulsar and/or that other sources, such as those discussed here, could contribute to the emission of the Milagro
source. The results presented in this paper may be useful in interpreting future data provided by the
{\it Fermi} satellite of the gamma-ray sources in this
remarkable region of the Galactic plane. The physical understanding of the most
relevant sources in the field is currently a work in progress, in addition to  the
analysis of new {\it XMM-Newton} and {\it AGILE} observations.

\begin{acknowledgements}

We thank the staff of the GMRT who have made these observations possible.
GMRT is run by the National Centre for Radio Astrophysics of the Tata Institute
of Fundamental Research. 
Based on observations collected at the Centro Astron\'omico Hispano Alem\'an
(CAHA) at Calar Alto, operated jointly by the Max-Planck Institut f\"ur
Astronomie and the Instituto de Astrof\'{\i}sica de Andaluc\'{\i}a (CSIC)
This publication makes use of data products from the Two Micron All Sky Survey,
which is a joint project of the University of Massachusetts and the Infrared
Processing and Analysis Center/California Institute of Technology, funded by
the NASA and NSF in the USA.
e-VLBI developments in Europe are supported by the EC DG-INFSO funded
Communication Network Developments project 'EXPReS' Contract No. 02662.
The European VLBI Network is a joint facility of European, Chinese, South
African and other radio astronomy institutes funded by their national research
councils.
We acknowledge support by DGI of the Spanish Ministerio de Educaci\'on y
Ci\-encia (MEC) under grants AYA2007-68034-C03-01, AYA2007-68034-C03-02 and
AYA2007-68034-C03-03, FEDER funds and Junta de Andaluc\'{\i}a under PAIDI
research group FQM-322. 
J.M. was supported by the Spanish Ministerio de Ciencia e Innovacion (MICINN)
under fellowship BES-2008-004564.
M.P. and M.R. acknowledge financial support from MEC and European Social Funds
through a \emph{Ram\'on y Cajal} research contract.
V.B-R. gratefully acknowledges support from the Alexander von Humboldt
Foundation.
P.B. was supported by the DGI of MEC (Spain) under fellowship BES-2005-7234.
G.E.R is supported by the Argentine Agencies CONICET (PIP 5375) and ANPCyT
(PICT 03-13291).
We thank the anonymous referee for his useful comments.

\end{acknowledgements}


\Online

{\scriptsize

\longtab{2}{
\begin{longtable}{ccccrcrccc}
\caption[]{List of GMRT sources detected at 610~MHz, including their names,
positions, peak flux densities, integrated flux densities, NIR magnitudes (dots
denote that the radio source is outside our NIR mosaic, while ND represents 
non-detections) and X-ray fluxes (dots imply that the radio source is outside
the X-ray fields, while ND stands for non-detections).}\\
\hline\hline
\#  & Name &  RA       & DEC       & Peak flux density & Local Noise       & Integrated flux density & Error & $K_{\rm s}$ & $F_{0.2-12~{\rm keV}}$ \\
    &      & (J2000.0) & (J2000.0) & (mJy~beam$^{-1}$) & (mJy~beam$^{-1}$) & (mJy) & (mJy)           & (mag) & (erg~cm$^{-2}$~s$^{-1}$) \\
\hline
  1 & GMRT~J201142.7+374208 & 20:11:42.73& +37:42:08.5 &    8.00 &  0.53 &    9.24 & 0.18 & ... & ND  \\
  2 & GMRT~J201146.3+364937 & 20:11:46.38& +36:49:37.5 &    4.59 &  0.33 &   24.81 & 0.30 & ... & ND  \\
  3 & GMRT~J201146.5+362243 & 20:11:46.58& +36:22:43.8 &   74.05 &  0.44 &  136.67 & 0.24 & ... & ... \\
  4 & GMRT~J201147.6+362234 & 20:11:47.62& +36:22:34.2 &   86.34 &  0.42 &  149.12 & 0.22 & ... & ... \\
  5 & GMRT~J201201.3+362753 & 20:12:01.33& +36:27:53.7 &   12.93 &  0.37 &   14.96 & 0.12 & ... & ... \\
  6 & GMRT~J201205.7+371130 & 20:12:05.71& +37:11:30.6 &    9.14 &  0.49 &   17.85 & 0.30 & ... & ... \\
  7 & GMRT~J201209.8+361841 & 20:12:09.85& +36:18:41.2 &    4.15 &  0.26 &    4.51 & 0.09 & ... & ... \\
  8 & GMRT~J201210.2+373305 & 20:12:10.20& +37:33:05.6 &    7.30 &  0.37 &    8.72 & 0.14 & ... & ... \\
  9 & GMRT~J201215.7+364050 & 20:12:15.74& +36:40:50.5 &    5.55 &  0.31 &    6.23 & 0.12 & ... & ... \\
 10 & GMRT~J201226.0+364915 & 20:12:26.09& +36:49:15.3 &    7.32 &  0.24 &    6.98 & 0.08 & ... & ... \\
 11 & GMRT~J201231.1+361933 & 20:12:31.14& +36:19:33.2 &   39.65 &  0.23 &   46.83 & 0.10 & ... & ... \\
 12 & GMRT~J201231.9+361939 & 20:12:31.98& +36:19:39.0 &   39.07 &  0.23 &   45.91 & 0.10 & ... & ... \\
 13 & GMRT~J201238.8+362608 & 20:12:38.86& +36:26:08.8 &    4.82 &  0.28 &    5.10 & 0.09 & ... & ... \\
 14 & GMRT~J201239.1+360441 & 20:12:39.18& +36:04:41.3 &    8.42 &  0.19 &    8.83 & 0.07 & ... & ... \\
 15 & GMRT~J201239.2+363457 & 20:12:39.24& +36:34:57.5 &   14.63 &  0.36 &   40.81 & 0.24 & ... & ND  \\
 16 & GMRT~J201239.6+372145 & 20:12:39.65& +37:21:45.7 &   11.85 &  0.40 &   17.96 & 0.18 & ... & ... \\
 17 & GMRT~J201239.7+362538 & 20:12:39.71& +36:25:38.8 &    5.06 &  0.26 &    5.22 & 0.08 & ... & ... \\
 18 & GMRT~J201240.2+363444 & 20:12:40.26& +36:34:44.0 &   16.26 &  0.36 &   36.98 & 0.22 & ... & ... \\
 19 & GMRT~J201242.7+365510 & 20:12:42.78& +36:55:10.4 &   11.65 &  0.24 &   20.67 & 0.11 & ... & ... \\
 20 & GMRT~J201243.7+374416 & 20:12:43.75& +37:44:16.5 &   46.67 &  0.38 &   54.87 & 0.16 & ... & ... \\
 21 & GMRT~J201245.3+363333 & 20:12:45.39& +36:33:33.1 &   41.41 &  0.33 &   65.06 & 0.16 & ... & ... \\
 22 & GMRT~J201246.6+361309 & 20:12:46.61& +36:13:09.7 &    4.06 &  0.23 &    4.07 & 0.08 & ... & ... \\
 23 & GMRT~J201248.2+374332 & 20:12:48.20& +37:43:32.5 &   18.02 &  0.35 &   19.03 & 0.12 & ... & ND  \\
 24 & GMRT~J201300.6+370004 & 20:13:00.65& +37:00:04.8 &    4.51 &  0.26 &    4.87 & 0.10 & ... & ... \\
 25 & GMRT~J201304.5+365736 & 20:13:04.57& +36:57:36.2 &   43.73 &  0.28 &   69.86 & 0.13 & ... & ... \\
 26 & GMRT~J201305.3+365739 & 20:13:05.35& +36:57:39.4 &   58.91 &  0.28 &  100.44 & 0.15 & ... & ... \\
 27 & GMRT~J201305.4+360134 & 20:13:05.41& +36:01:34.5 &   21.11 &  0.22 &   21.95 & 0.09 & ... & ND  \\
 28 & GMRT~J201317.0+370715 & 20:13:17.03& +37:07:15.8 &   14.39 &  0.29 &   16.44 & 0.11 & ... & ND  \\
 29 & GMRT~J201319.6+373729 & 20:13:19.66& +37:37:29.4 &    8.81 &  0.30 &    9.25 & 0.12 & ... & ... \\
 30 & GMRT~J201320.9+375132 & 20:13:20.94& +37:51:32.4 &   30.36 &  0.40 &   37.25 & 0.16 & ... & ... \\
 31 & GMRT~J201324.1+375516 & 20:13:24.13& +37:55:16.7 &   10.60 &  0.45 &   13.89 & 0.17 & ... & ... \\
 32 & GMRT~J201334.0+361501 & 20:13:34.03& +36:15:01.8 &   35.94 &  0.26 &   82.97 & 0.15 & ... & ... \\
 33 & GMRT~J201334.3+360926 & 20:13:34.33& +36:09:26.0 &  224.70 &  0.37 &  376.50 & 0.22 & ... & ... \\
 34 & GMRT~J201337.0+360942 & 20:13:37.01& +36:09:42.7 &   54.91 &  0.38 &  241.40 & 0.30 & ... & ... \\
 35 & GMRT~J201346.1+365908 & 20:13:46.19& +36:59:08.2 &    5.20 &  0.24 &    5.45 & 0.08 & ... & ... \\
 36 & GMRT~J201347.5+365539 & 20:13:47.56& +36:55:39.4 &    3.49 &  0.23 &    5.49 & 0.10 & ... & ND  \\
 37 & GMRT~J201347.7+373920 & 20:13:47.70& +37:39:20.5 &    9.85 &  0.33 &   13.32 & 0.14 & ... & ND  \\
 38 & GMRT~J201349.1+355827 & 20:13:49.11& +35:58:27.9 &    6.72 &  0.26 &    5.90 & 0.07 & ... & ... \\
 39 & GMRT~J201405.5+372431 & 20:14:05.59& +37:24:31.4 &   12.58 &  0.38 &   15.84 & 0.15 & ... & ND  \\
 40 & GMRT~J201408.7+373325 & 20:14:08.77& +37:33:25.4 &   28.25 &  0.37 &   47.09 & 0.18 & ... & ... \\
 41 & GMRT~J201409.4+373400 & 20:14:09.49& +37:34:00.1 &   50.56 &  0.39 &   74.39 & 0.19 & ... & ... \\
 42 & GMRT~J201410.4+371552 & 20:14:10.48& +37:15:52.8 &   23.05 &  0.50 &   36.08 & 0.23 & ... & ... \\
 43 & GMRT~J201410.7+371544 & 20:14:10.71& +37:15:44.1 &   14.84 &  0.50 &   21.21 & 0.21 & ... & ... \\
 44 & GMRT~J201412.4+355218 & 20:14:12.48& +35:52:18.0 &    8.82 &  0.30 &   16.61 & 0.15 & ... & ... \\
 45 & GMRT~J201413.4+355242 & 20:14:13.45& +35:52:42.7 &    5.31 &  0.29 &    8.41 & 0.13 & ... & ... \\
 46 & GMRT~J201416.1+372344 & 20:14:16.16& +37:23:44.0 &    7.36 &  0.42 &   28.14 & 0.30 & ... & ... \\
 47 & GMRT~J201418.2+372339 & 20:14:18.23& +37:23:39.7 &    5.58 &  0.44 &   13.50 & 0.19 & ... & ND  \\
 48 & GMRT~J201425.7+353650 & 20:14:25.78& +35:36:50.1 &   14.38 &  0.34 &   24.24 & 0.16 & ... & ... \\
 49 & GMRT~J201435.8+364550 & 20:14:35.80& +36:45:50.7 &    4.49 &  0.34 &   13.36 & 0.18 & ... & ... \\
 50 & GMRT~J201449.4+374335 & 20:14:49.45& +37:43:35.1 &   38.17 &  0.34 &   75.05 & 0.19 & ... & ... \\
 51 & GMRT~J201450.9+360136 & 20:14:50.94& +36:01:36.6 &   95.37 &  0.39 &  264.77 & 0.30 & ... & ... \\
 52 & GMRT~J201451.6+360149 & 20:14:51.64& +36:01:49.9 &   31.94 &  0.39 &  138.58 & 0.30 & ... & ... \\
 53 & GMRT~J201451.8+370025 & 20:14:51.82& +37:00:25.6 &   14.90 &  0.56 &   23.33 & 0.24 & ... & ... \\
 54 & GMRT~J201452.0+361758 & 20:14:52.07& +36:17:58.6 &   13.67 &  0.33 &   17.10 & 0.12 & ... & ... \\
 55 & GMRT~J201454.4+370034 & 20:14:54.46& +37:00:34.6 &   16.24 &  0.56 &   25.47 & 0.21 & ... & ... \\
 56 & GMRT~J201509.0+373725 & 20:15:09.05& +37:37:25.0 &    7.66 &  0.33 &    9.47 & 0.12 & ... & ... \\
 57 & GMRT~J201509.3+371655 & 20:15:09.32& +37:16:55.6 &   17.09 &  0.88 &   60.91 & 0.60 & ... & ... \\
 58 & GMRT~J201510.6+370049 & 20:15:10.69& +37:00:49.5 &   10.73 &  0.54 &   21.55 & 0.30 & ... & ... \\
 59 & GMRT~J201516.5+362701 & 20:15:16.52& +36:27:01.5 &  401.95 &  0.69 &  632.20 & 0.50 & ... & ... \\
 60 & GMRT~J201517.1+362705 & 20:15:17.17& +36:27:05.3 &  294.24 &  0.69 &  415.41 & 0.40 & ... & ND  \\
 61 & GMRT~J201518.9+353616 & 20:15:18.91& +35:36:16.1 &   10.61 &  0.22 &   15.70 & 0.09 & ... & ... \\
 62 & GMRT~J201527.4+353621 & 20:15:27.47& +35:36:21.1 &   22.28 &  0.25 &   37.70 & 0.12 & ... & ... \\
 63 & GMRT~J201528.7+371100 & 20:15:28.77& +37:11:00.2 & 1309.26 &  1.26 & 1891.45 & 0.70 & ... & ... \\
 64 & GMRT~J201529.7+380119 & 20:15:29.70& +38:01:19.0 &   49.27 &  0.29 &   74.70 & 0.15 & ... & ... \\
 65 & GMRT~J201529.7+363943 & 20:15:29.71& +36:39:43.5 &    5.07 &  0.31 &    6.63 & 0.11 & ... & ND  \\
 66 & GMRT~J201534.7+375728 & 20:15:34.76& +37:57:28.7 &    8.19 &  0.20 &   29.68 & 0.17 & ... & ND  \\
 67 & GMRT~J201535.1+363928 & 20:15:35.15& +36:39:28.4 &    3.96 &  0.30 &    7.07 & 0.13 & ... & ... \\
 68 & GMRT~J201535.5+375718 & 20:15:35.57& +37:57:18.4 &    7.11 &  0.20 &   50.31 & 0.30 & ... & ... \\
 69 & GMRT~J201556.3+365935 & 20:15:56.35& +36:59:35.6 &   80.35 &  0.40 &  103.09 & 0.18 & ... & ... \\
 70 & GMRT~J201557.8+375014 & 20:15:57.80& +37:50:14.7 &    7.19 &  0.17 &    8.70 & 0.06 & ... & ... \\
 71 & GMRT~J201558.1+381003 & 20:15:58.12& +38:10:03.6 &    7.62 &  0.23 &   10.78 & 0.10 & ... & $4.5\pm0.1\times10^{-13}$ \\
 72 & GMRT~J201558.5+362559 & 20:15:58.56& +36:25:59.1 &   31.57 &  0.26 &   73.82 & 0.17 & ... & ... \\
 73 & GMRT~J201558.7+381004 & 20:15:58.73& +38:10:04.1 &   12.22 &  0.23 &   23.42 & 0.13 & ... & ... \\
 74 & GMRT~J201559.9+362536 & 20:15:59.90& +36:25:36.8 &    6.07 &  0.26 &   10.12 & 0.15 & ... & ... \\
 75 & GMRT~J201600.7+362517 & 20:16:00.71& +36:25:17.1 &   13.06 &  0.26 &   54.98 & 0.23 & ... & ... \\
 76 & GMRT~J201603.2+375721 & 20:16:03.20& +37:57:21.6 &   39.66 &  0.19 &   58.18 & 0.10 & ... & ... \\
 77 & GMRT~J201616.9+353948 & 20:16:16.95& +35:39:48.4 &   50.40 &  0.29 &  180.26 & 0.23 & ... & ... \\
 78 & GMRT~J201619.8+380044 & 20:16:19.82& +38:00:44.2 &    3.38 &  0.16 &    4.39 & 0.07 & ... & ... \\
 79 & GMRT~J201620.9+353945 & 20:16:20.90& +35:39:45.7 &   97.32 &  0.29 &  314.61 & 0.21 & ... & ... \\
 80 & GMRT~J201621.4+354020 & 20:16:21.41& +35:40:20.6 &    4.15 &  0.28 &    7.07 & 0.12 & ... & ... \\
 81 & GMRT~J201621.6+380519 & 20:16:21.60& +38:05:19.9 &    2.15 &  0.18 &    3.48 & 0.07 & ... & ... \\
 82 & GMRT~J201626.0+355829 & 20:16:26.03& +35:58:29.5 &    5.76 &  0.28 &    8.36 & 0.12 & ... & ... \\
 83 & GMRT~J201627.5+365501 & 20:16:27.51& +36:55:01.1 &   19.55 &  0.26 &   38.45 & 0.13 & ... & ... \\
 84 & GMRT~J201641.9+353650 & 20:16:41.98& +35:36:50.3 &   48.59 &  0.28 &   88.20 & 0.14 & ... & ... \\
 85 & GMRT~J201645.3+360034 & 20:16:45.38& +36:00:34.8 &  325.22 &  0.55 &  698.65 & 0.40 & ... & ... \\
 86 & GMRT~J201645.6+360109 & 20:16:45.66& +36:01:09.7 &   15.07 &  0.58 &   20.37 & 0.23 & ... & ... \\
 87 & GMRT~J201646.2+380601 & 20:16:46.26& +38:06:01.9 &    2.80 &  0.17 &    3.35 & 0.07 & ... & ... \\
 88 & GMRT~J201648.2+363133 & 20:16:48.23& +36:31:33.8 &   13.05 &  0.18 &   44.51 & 0.13 & ... & ND  \\
 89 & GMRT~J201654.2+372553 & 20:16:54.27& +37:25:53.0 &   15.98 &  0.41 &   24.75 & 0.23 & ... & ... \\
 90 & GMRT~J201654.4+363104 & 20:16:54.46& +36:31:04.5 &    4.44 &  0.18 &    9.43 & 0.11 & ... & ... \\
 91 & GMRT~J201656.4+371353 & 20:16:56.46& +37:13:53.8 &    3.98 &  0.25 &    4.93 & 0.10 & ... & ... \\
 92 & GMRT~J201657.6+370545 & 20:16:57.62& +37:05:45.4 &    6.43 &  0.21 &   11.75 & 0.09 & ... & ... \\
 93 & GMRT~J201659.9+363231 & 20:16:59.99& +36:32:31.0 &    2.76 &  0.18 &    2.57 & 0.06 & ... & ... \\
 94 & GMRT~J201700.5+360126 & 20:17:00.52& +36:01:26.3 &    9.45 &  0.42 &   14.53 & 0.15 & ... & ... \\
 95 & GMRT~J201700.5+354712 & 20:17:00.55& +35:47:12.8 &   11.58 &  0.27 &   23.36 & 0.14 & ... & ... \\
 96 & GMRT~J201701.5+354654 & 20:17:01.53& +35:46:54.3 &   39.87 &  0.33 &   87.71 & 0.19 & ... & ... \\
 97 & GMRT~J201716.8+375819 & 20:17:16.83& +37:58:19.3 &   12.53 &  0.17 &   23.37 & 0.09 & ... & ... \\
 98 & GMRT~J201720.8+350749 & 20:17:20.88& +35:07:49.9 &    5.90 &  0.34 &   10.16 & 0.15 & ... & ... \\
 99 & GMRT~J201721.9+354610 & 20:17:21.98& +35:46:10.7 &   17.31 &  0.30 &   47.70 & 0.17 & ... & ... \\
100 & GMRT~J201725.8+373043 & 20:17:25.88& +37:30:43.1 &    6.84 &  0.41 &   11.19 & 0.18 & ... & ... \\
101 & GMRT~J201726.7+371357 & 20:17:26.70& +37:13:57.2 &    9.57 &  0.19 &   15.02 & 0.10 & ND  & ... \\
102 & GMRT~J201727.9+355144 & 20:17:27.97& +35:51:44.1 &   12.91 &  0.26 &   21.93 & 0.12 & ... & ... \\
103 & GMRT~J201732.1+371605 & 20:17:32.15& +37:16:05.9 &    3.09 &  0.19 &    3.30 & 0.07 & ND  & ... \\
104 & GMRT~J201741.4+355629 & 20:17:41.42& +35:56:29.3 &    4.63 &  0.26 &   12.86 & 0.16 & ... & ... \\
105 & GMRT~J201742.1+355628 & 20:17:42.10& +35:56:28.4 &    3.89 &  0.26 &   10.35 & 0.15 & ... & ... \\
106 & GMRT~J201742.1+373507 & 20:17:42.15& +37:35:07.6 &    9.83 &  0.31 &   14.15 & 0.15 & ... & ... \\
107 & GMRT~J201742.5+372501 & 20:17:42.55& +37:25:01.8 &    4.92 &  0.25 &    5.06 & 0.07 & ... & ... \\
108 & GMRT~J201744.3+365142 & 20:17:44.37& +36:51:42.2 &    5.89 &  0.16 &   15.00 & 0.13 & 16.0 & ND \\
109 & GMRT~J201744.8+365045 & 20:17:44.82& +36:50:45.2 &   22.12 &  0.17 &   33.46 & 0.08 & ND & $5.5\pm0.5\times10^{-13}$ \\
110 & GMRT~J201745.4+365315 & 20:17:45.48& +36:53:15.6 &    8.61 &  0.17 &   14.93 & 0.09 & 14.9 & ND \\
111 & GMRT~J201748.5+371322 & 20:17:48.58& +37:13:22.2 &    3.08 &  0.18 &    3.57 & 0.07 & ND  & ... \\
112 & GMRT~J201748.6+351833 & 20:17:48.60& +35:18:33.9 &    4.10 &  0.30 &    5.43 & 0.11 & ... & ... \\
113 & GMRT~J201748.7+375807 & 20:17:48.76& +37:58:07.0 &    4.36 &  0.19 &    7.02 & 0.09 & ... & ... \\
114 & GMRT~J201749.5+381549 & 20:17:49.59& +38:15:49.6 &  206.04 &  0.48 &  417.19 & 0.30 & ... & ... \\
115 & GMRT~J201749.9+365508 & 20:17:49.90& +36:55:08.3 &    5.11 &  0.18 &    6.57 & 0.08 & 16.6 & $1.0\pm0.3\times10^{-13}$ \\
116 & GMRT~J201750.4+370311 & 20:17:50.48& +37:03:11.3 &    4.67 &  0.16 &    5.18 & 0.06 & ND  & ... \\
117 & GMRT~J201750.5+353407 & 20:17:50.52& +35:34:07.9 &    3.11 &  0.25 &    3.04 & 0.07 & ... & ... \\
118 & GMRT~J201751.5+351821 & 20:17:51.51& +35:18:21.6 &   77.25 &  0.29 &  122.32 & 0.16 & ... & ... \\
119 & GMRT~J201753.7+381522 & 20:17:53.73& +38:15:22.8 &    7.03 &  0.46 &   10.21 & 0.18 & ... & ... \\
120 & GMRT~J201753.8+351822 & 20:17:53.88& +35:18:22.2 &  104.31 &  0.27 &  189.99 & 0.17 & ... & ... \\
121 & GMRT~J201754.8+375451 & 20:17:54.84& +37:54:51.5 &   20.96 &  0.22 &   26.22 & 0.09 & ... & ND  \\
122 & GMRT~J201755.1+380017 & 20:17:55.10& +38:00:17.5 &    3.13 &  0.20 &    5.01 & 0.09 & ... & ND  \\
123 & GMRT~J201756.3+363726 & 20:17:56.35& +36:37:26.9 &   11.41 &  0.25 &   16.38 & 0.11 & ND  & ND  \\
124 & GMRT~J201756.5+364540 & 20:17:56.56& +36:45:40.9 &   15.95 &  0.20 &   51.76 & 0.13 & ND & $13.4\pm0.7\times10^{-13}$\\
125 & GMRT~J201756.5+364825 & 20:17:56.59& +36:48:25.3 &   12.40 &  0.16 &   16.51 & 0.08 & ND  & ... \\
126 & GMRT~J201758.2+370125 & 20:17:58.29& +37:01:25.3 &    2.28 &  0.17 &    2.57 & 0.06 & 13.5 & ... \\
127 & GMRT~J201758.8+351806 & 20:17:58.83& +35:18:06.3 &    3.16 &  0.25 &    6.00 & 0.11 & ... & ... \\
128 & GMRT~J201759.7+353655 & 20:17:59.70& +35:36:55.5 &   14.47 &  0.28 &   25.90 & 0.14 & ... & ... \\
129 & GMRT~J201759.7+363018 & 20:17:59.75& +36:30:18.1 &   29.44 &  0.21 &   39.47 & 0.09 & ND  & ... \\
130 & GMRT~J201803.9+375314 & 20:18:03.91& +37:53:14.6 &    4.45 &  0.22 &    5.55 & 0.09 & ... & ... \\
131 & GMRT~J201807.4+381436 & 20:18:07.46& +38:14:36.1 &    7.84 &  0.30 &    9.09 & 0.10 & ... & ... \\
132 & GMRT~J201807.6+360357 & 20:18:07.64& +36:03:57.2 &    7.70 &  0.21 &   13.11 & 0.09 & ... & ... \\
133 & GMRT~J201808.0+345816 & 20:18:08.04& +34:58:16.2 &   12.21 &  0.39 &   27.64 & 0.19 & ... & ... \\
134 & GMRT~J201810.1+371936 & 20:18:10.16& +37:19:36.5 &    3.11 &  0.17 &    8.97 & 0.08 & ... & ... \\
135 & GMRT~J201811.2+354551 & 20:18:11.25& +35:45:51.1 &    2.68 &  0.23 &    3.77 & 0.07 & ... & ... \\
136 & GMRT~J201812.7+350746 & 20:18:12.79& +35:07:46.7 &    5.00 &  0.21 &    5.35 & 0.06 & ... & ... \\
137 & GMRT~J201813.0+360046 & 20:18:13.08& +36:00:46.3 &   14.58 &  0.21 &   19.16 & 0.10 & ... & ... \\
138 & GMRT~J201815.4+352032 & 20:18:15.45& +35:20:32.5 &   10.38 &  0.18 &   11.18 & 0.07 & ... & ... \\
139 & GMRT~J201822.6+351212 & 20:18:22.61& +35:12:12.8 &    4.80 &  0.19 &    5.67 & 0.08 & ... & ND  \\
140 & GMRT~J201823.9+370819 & 20:18:23.94& +37:08:19.4 &   27.63 &  0.16 &   37.99 & 0.08 & ND  & ... \\
141 & GMRT~J201830.6+370225 & 20:18:30.64& +37:02:25.6 &    7.55 &  0.17 &   61.16 & 0.20 & ND  & ... \\
142 & GMRT~J201832.4+370234 & 20:18:32.43& +37:02:34.5 &   13.66 &  0.16 &  103.51 & 0.21 & ND  & ... \\
143 & GMRT~J201833.5+374016 & 20:18:33.53& +37:40:16.7 &    3.72 &  0.22 &   10.24 & 0.13 & ... & ... \\
144 & GMRT~J201834.4+374020 & 20:18:34.45& +37:40:20.8 &    9.48 &  0.22 &   17.34 & 0.13 & ... & ... \\
145 & GMRT~J201839.3+380853 & 20:18:39.32& +38:08:53.6 &   77.29 &  0.30 &  269.62 & 0.21 & ... & ... \\
146 & GMRT~J201841.8+361717 & 20:18:41.86& +36:17:17.8 &   10.01 &  0.25 &   14.25 & 0.11 & ... & ... \\
147 & GMRT~J201842.7+381242 & 20:18:42.79& +38:12:42.0 &   72.15 &  0.25 &   84.24 & 0.11 & ... & ... \\
148 & GMRT~J201843.7+351006 & 20:18:43.78& +35:10:06.2 &    2.39 &  0.18 &    2.44 & 0.06 & ... & ... \\
149 & GMRT~J201843.7+381656 & 20:18:43.79& +38:16:56.6 &    4.28 &  0.21 &    4.23 & 0.07 & ... & ... \\
150 & GMRT~J201844.9+375108 & 20:18:44.93& +37:51:08.8 &    5.09 &  0.21 &    6.35 & 0.08 & ... & ... \\
151 & GMRT~J201848.4+353512 & 20:18:48.43& +35:35:12.9 &   11.20 &  0.18 &   12.45 & 0.07 & ... & ND  \\
152 & GMRT~J201848.4+353137 & 20:18:48.49& +35:31:37.2 &    4.76 &  0.16 &    7.74 & 0.09 & ... & ... \\
153 & GMRT~J201848.5+371937 & 20:18:48.50& +37:19:37.7 &    6.76 &  0.16 &    8.57 & 0.08 & ... & ... \\
154 & GMRT~J201850.2+351734 & 20:18:50.24& +35:17:34.7 &    3.37 &  0.15 &    3.64 & 0.05 & ... & ... \\
155 & GMRT~J201850.8+350926 & 20:18:50.80& +35:09:26.0 &    5.68 &  0.18 &   13.84 & 0.13 & ... & ... \\
156 & GMRT~J201851.6+350930 & 20:18:51.60& +35:09:30.0 &    4.02 &  0.18 &    8.61 & 0.10 & ... & ... \\
157 & GMRT~J201852.4+363632 & 20:18:52.46& +36:36:32.0 &   32.61 &  0.22 &   64.00 & 0.12 & ND  & ... \\
158 & GMRT~J201854.0+355023 & 20:18:54.07& +35:50:23.1 &    3.83 &  0.23 &    3.56 & 0.07 & ... & ND  \\
159 & GMRT~J201854.6+352821 & 20:18:54.64& +35:28:21.4 &    5.39 &  0.16 &    5.26 & 0.05 & ... & ... \\
160 & GMRT~J201855.5+360213 & 20:18:55.50& +36:02:13.1 &    2.54 &  0.17 &    2.38 & 0.06 & ... & ... \\
161 & GMRT~J201856.5+360609 & 20:18:56.58& +36:06:09.9 &    2.81 &  0.19 &    3.08 & 0.07 & ... & ... \\
162 & GMRT~J201858.0+375656 & 20:18:58.02& +37:56:56.6 &    2.82 &  0.23 &    7.95 & 0.12 & ... & ... \\
163 & GMRT~J201859.5+372451 & 20:18:59.58& +37:24:51.2 &   11.17 &  0.21 &   16.11 & 0.09 & ... & ... \\
164 & GMRT~J201904.0+370206 & 20:19:04.04& +37:02:06.6 &    3.17 &  0.12 &    4.57 & 0.05 & ND  & ... \\
165 & GMRT~J201904.8+360809 & 20:19:04.83& +36:08:09.8 &    4.86 &  0.20 &    4.52 & 0.07 & ... & ... \\
166 & GMRT~J201908.7+374925 & 20:19:08.78& +37:49:25.8 &   11.24 &  0.24 &   13.52 & 0.10 & ... & ... \\
167 & GMRT~J201913.7+352154 & 20:19:13.72& +35:21:54.3 &   49.36 &  0.20 &   60.42 & 0.10 & ... & ... \\
168 & GMRT~J201914.5+351518 & 20:19:14.56& +35:15:18.7 &    2.28 &  0.16 &    4.01 & 0.07 & ... & ND  \\
169 & GMRT~J201914.5+354400 & 20:19:14.58& +35:44:00.7 &   58.11 &  0.25 &   90.26 & 0.13 & ... & ... \\
170 & GMRT~J201914.6+352156 & 20:19:14.67& +35:21:56.9 &   28.84 &  0.20 &   43.13 & 0.10 & ... & ... \\
171 & GMRT~J201916.0+373528 & 20:19:16.03& +37:35:28.5 &    9.35 &  0.16 &   12.51 & 0.07 & ... & ... \\
172 & GMRT~J201916.6+371151 & 20:19:16.64& +37:11:51.0 &    2.35 &  0.12 &   12.85 & 0.11 & ND  & ... \\
173 & GMRT~J201917.5+373553 & 20:19:17.51& +37:35:53.9 &    9.41 &  0.17 &    9.13 & 0.06 & ... & ... \\
174 & GMRT~J201918.2+355025 & 20:19:18.25& +35:50:25.1 &   18.46 &  0.23 &   41.69 & 0.12 & ... & ... \\
175 & GMRT~J201918.3+373828 & 20:19:18.35& +37:38:28.6 &   23.21 &  0.16 &   25.67 & 0.07 & ... & ... \\
176 & GMRT~J201920.0+363750 & 20:19:20.04& +36:37:50.5 &   20.59 &  0.22 &   30.44 & 0.10 & 16.9 & ... \\
177 & GMRT~J201920.4+380314 & 20:19:20.48& +38:03:14.7 &    3.03 &  0.17 &    3.04 & 0.06 & ... & ... \\
178 & GMRT~J201920.9+373504 & 20:19:20.95& +37:35:04.2 &    2.16 &  0.17 &    4.02 & 0.08 & ... & ... \\
179 & GMRT~J201922.0+352227 & 20:19:22.09& +35:22:27.0 &    2.17 &  0.18 &    2.39 & 0.06 & ... & ND  \\
180 & GMRT~J201925.9+354158 & 20:19:25.92& +35:41:58.9 &   34.62 &  0.25 &   50.12 & 0.12 & ... & ... \\
181 & GMRT~J201926.9+381656 & 20:19:26.91& +38:16:56.1 &    6.35 &  0.16 &    6.30 & 0.06 & ... & ... \\
182 & GMRT~J201928.1+362610 & 20:19:28.11& +36:26:10.4 &   12.91 &  0.19 &   28.06 & 0.10 & 15.6 & ... \\
183 & GMRT~J201930.6+375339 & 20:19:30.62& +37:53:39.8 &    3.23 &  0.20 &    4.16 & 0.08 & ... & ... \\
184 & GMRT~J201931.8+372423 & 20:19:31.81& +37:24:23.1 &    4.03 &  0.16 &   12.35 & 0.11 & ... & ... \\
185 & GMRT~J201932.3+372440 & 20:19:32.31& +37:24:40.3 &    3.10 &  0.17 &    9.80 & 0.11 & ... & ND  \\
186 & GMRT~J201933.9+370440 & 20:19:33.95& +37:04:40.2 &   34.03 &  0.14 &   56.00 & 0.07 & ND  & ... \\
187 & GMRT~J201940.6+350424 & 20:19:40.67& +35:04:24.4 &    2.97 &  0.25 &    3.35 & 0.08 & ... & ... \\
188 & GMRT~J201941.2+361144 & 20:19:41.24& +36:11:44.1 &   11.58 &  0.26 &   13.62 & 0.10 & ... & ... \\
189 & GMRT~J201943.2+372956 & 20:19:43.24& +37:29:56.0 &   17.74 &  0.24 &   24.76 & 0.10 & ... & ... \\
190 & GMRT~J201943.7+353224 & 20:19:43.70& +35:32:24.3 &    7.89 &  0.19 &   13.47 & 0.08 & ... & ... \\
191 & GMRT~J201943.9+371909 & 20:19:43.93& +37:19:09.7 &    4.40 &  0.14 &    5.19 & 0.07 & ... & ... \\
192 & GMRT~J201945.4+351826 & 20:19:45.41& +35:18:26.8 &    3.13 &  0.17 &    2.99 & 0.06 & ... & ... \\
193 & GMRT~J201947.4+370634 & 20:19:47.47& +37:06:34.3 &    3.70 &  0.16 &   10.88 & 0.10 & ND  & ... \\
194 & GMRT~J201948.1+370645 & 20:19:48.12& +37:06:45.6 &    6.85 &  0.16 &   17.08 & 0.11 & ND  & ND  \\
195 & GMRT~J201950.2+382949 & 20:19:50.22& +38:29:49.5 &   16.78 &  0.19 &   19.26 & 0.07 & ... & ... \\
196 & GMRT~J201951.1+362936 & 20:19:51.13& +36:29:36.3 &   34.80 &  0.25 &   53.50 & 0.12 & ND  & ... \\
197 & GMRT~J201952.4+354727 & 20:19:52.40& +35:47:27.4 &    5.07 &  0.24 &    6.44 & 0.09 & ... & ... \\
198 & GMRT~J201953.8+350353 & 20:19:53.87& +35:03:53.6 &    3.58 &  0.27 &    7.45 & 0.12 & ... & ... \\
199 & GMRT~J201955.3+371757 & 20:19:55.37& +37:17:57.3 &    7.78 &  0.13 &    9.19 & 0.06 & ND  & ... \\
200 & GMRT~J201956.8+373914 & 20:19:56.84& +37:39:14.3 &    3.51 &  0.16 &    2.94 & 0.05 & ... & ... \\
201 & GMRT~J201958.7+381427 & 20:19:58.78& +38:14:27.4 &    2.58 &  0.15 &    2.36 & 0.05 & ... & ... \\
202 & GMRT~J201959.2+371833 & 20:19:59.25& +37:18:33.0 &    4.93 &  0.15 &   10.18 & 0.09 & ... & ND  \\
203 & GMRT~J202000.5+365806 & 20:20:00.52& +36:58:06.7 &    2.10 &  0.13 &    2.71 & 0.05 & ND  & ... \\
204 & GMRT~J202000.7+351809 & 20:20:00.74& +35:18:09.7 &   17.71 &  0.22 &   69.92 & 0.20 & ... & ND  \\
205 & GMRT~J202000.9+351829 & 20:20:00.99& +35:18:29.1 &    3.16 &  0.23 &   21.59 & 0.19 & ... & ... \\
206 & GMRT~J202001.5+351736 & 20:20:01.51& +35:17:36.3 &    9.55 &  0.21 &   49.79 & 0.20 & ... & ... \\
207 & GMRT~J202003.7+375018 & 20:20:03.75& +37:50:18.0 &   54.54 &  0.23 &   73.58 & 0.11 & ... & ... \\
208 & GMRT~J202003.9+373135 & 20:20:03.92& +37:31:35.8 &    8.21 &  0.30 &   17.49 & 0.17 & ... & ... \\
209 & GMRT~J202007.5+352415 & 20:20:07.57& +35:24:15.5 &   22.37 &  0.23 &  127.83 & 0.22 & ... & ... \\
210 & GMRT~J202008.0+374027 & 20:20:08.04& +37:40:27.8 &    4.73 &  0.18 &    4.35 & 0.07 & ... & ... \\
211 & GMRT~J202008.4+370147 & 20:20:08.47& +37:01:47.8 &    4.59 &  0.12 &    6.91 & 0.07 & ND  & ... \\
212 & GMRT~J202010.5+365749 & 20:20:10.56& +36:57:49.3 &    2.68 &  0.14 &    2.63 & 0.05 & ND  & ... \\
213 & GMRT~J202011.1+362246 & 20:20:11.16& +36:22:46.1 &    2.98 &  0.22 &    3.96 & 0.08 & ... & ... \\
214 & GMRT~J202011.6+354916 & 20:20:11.65& +35:49:16.4 &    3.92 &  0.23 &    5.66 & 0.10 & ... & ND  \\
215 & GMRT~J202011.9+362335 & 20:20:11.97& +36:23:35.7 &   12.46 &  0.22 &   22.50 & 0.11 & ... & ... \\
216 & GMRT~J202012.6+374016 & 20:20:12.67& +37:40:16.2 &    2.94 &  0.18 &    3.61 & 0.07 & ... & ... \\
217 & GMRT~J202020.2+370059 & 20:20:20.22& +37:00:59.7 &    1.65 &  0.12 &    1.69 & 0.04 & ND  & ... \\
218 & GMRT~J202020.6+382845 & 20:20:20.64& +38:28:45.0 &    3.21 &  0.18 &    3.24 & 0.06 & ... & ... \\
219 & GMRT~J202022.0+352459 & 20:20:22.09& +35:24:59.9 &    8.46 &  0.26 &   20.32 & 0.15 & ... & ... \\
220 & GMRT~J202022.1+372843 & 20:20:22.11& +37:28:43.0 &   62.30 &  0.26 &   88.20 & 0.13 & ... & ... \\
221 & GMRT~J202026.0+360726 & 20:20:26.07& +36:07:26.7 &   35.89 &  0.35 &   88.69 & 0.21 & ... & ND  \\
222 & GMRT~J202029.1+364212 & 20:20:29.10& +36:42:12.5 &    3.60 &  0.20 &    5.08 & 0.08 & ND  & ... \\
223 & GMRT~J202029.6+355131 & 20:20:29.65& +35:51:31.8 &   11.72 &  0.21 &   19.07 & 0.10 & ... & ... \\
224 & GMRT~J202029.8+353821 & 20:20:29.81& +35:38:21.1 &    5.88 &  0.22 &    8.48 & 0.10 & ... & ... \\
225 & GMRT~J202033.0+363159 & 20:20:33.03& +36:31:59.5 &  421.37 &  0.43 &  833.51 & 0.30 & ND  & ... \\
226 & GMRT~J202035.3+363130 & 20:20:35.32& +36:31:30.2 &    5.65 &  0.44 &    8.35 & 0.16 & ND  & ND  \\
227 & GMRT~J202036.4+373634 & 20:20:36.43& +37:36:34.9 &    8.22 &  0.17 &   10.24 & 0.10 & ... & ... \\
228 & GMRT~J202038.6+364721 & 20:20:38.65& +36:47:21.1 &    4.63 &  0.18 &    8.42 & 0.08 & ND  & ... \\
229 & GMRT~J202039.6+352621 & 20:20:39.64& +35:26:21.0 &   58.18 &  0.27 &   95.75 & 0.14 & ... & ... \\
230 & GMRT~J202040.8+362113 & 20:20:40.87& +36:21:13.1 &    8.80 &  0.20 &   16.28 & 0.09 & ... & ... \\
231 & GMRT~J202043.9+381811 & 20:20:43.91& +38:18:11.4 &   17.22 &  0.15 &   26.29 & 0.08 & ... & ... \\
232 & GMRT~J202044.8+354540 & 20:20:44.87& +35:45:40.9 &    4.10 &  0.25 &   10.66 & 0.14 & ... & ... \\
233 & GMRT~J202046.7+362731 & 20:20:46.78& +36:27:31.8 &    4.25 &  0.33 &    4.60 & 0.10 & ND  & ... \\
234 & GMRT~J202046.7+370650 & 20:20:46.79& +37:06:50.0 &   20.31 &  0.17 &   27.76 & 0.08 & ND  & ... \\
235 & GMRT~J202047.6+351429 & 20:20:47.60& +35:14:29.1 &    7.24 &  0.35 &   26.85 & 0.30 & ... & ... \\
236 & GMRT~J202047.8+353249 & 20:20:47.85& +35:32:49.8 &   10.52 &  0.23 &   14.88 & 0.11 & ... & ... \\
237 & GMRT~J202048.3+380858 & 20:20:48.31& +38:08:58.0 &   45.50 &  0.18 &   67.08 & 0.10 & ... & ... \\
238 & GMRT~J202054.3+371120 & 20:20:54.39& +37:11:20.5 &    7.91 &  0.15 &   12.65 & 0.08 & ND  & ... \\
239 & GMRT~J202055.8+383625 & 20:20:55.89& +38:36:25.1 &   37.89 &  0.35 &   49.79 & 0.16 & ... & ... \\
240 & GMRT~J202101.6+354028 & 20:21:01.66& +35:40:28.7 &    3.75 &  0.19 &    4.20 & 0.07 & ... & ... \\
241 & GMRT~J202105.3+382326 & 20:21:05.39& +38:23:26.0 &    6.71 &  0.19 &    6.92 & 0.06 & ... & ... \\
242 & GMRT~J202106.6+352441 & 20:21:06.69& +35:24:41.3 &   10.14 &  0.26 &   17.39 & 0.14 & ... & ... \\
243 & GMRT~J202109.3+371446 & 20:21:09.34& +37:14:46.1 &    3.49 &  0.19 &    4.40 & 0.06 & ND  & ... \\
244 & GMRT~J202116.6+362128 & 20:21:16.68& +36:21:28.5 &    5.33 &  0.19 &    6.22 & 0.08 & ... & ... \\
245 & GMRT~J202120.1+362229 & 20:21:20.12& +36:22:29.1 &    2.66 &  0.20 &    2.39 & 0.06 & ... & ... \\
246 & GMRT~J202120.2+352628 & 20:21:20.24& +35:26:28.1 &   86.25 &  0.24 &  113.62 & 0.12 & ... & ... \\
247 & GMRT~J202121.7+355226 & 20:21:21.75& +35:52:26.4 &    4.01 &  0.26 &    8.78 & 0.13 & ... & ... \\
248 & GMRT~J202126.9+382542 & 20:21:26.97& +38:25:42.9 &   11.76 &  0.24 &   13.77 & 0.09 & ... & ... \\
249 & GMRT~J202128.6+354623 & 20:21:28.62& +35:46:23.1 &    5.33 &  0.20 &    5.90 & 0.08 & ... & ... \\
250 & GMRT~J202131.0+355452 & 20:21:31.03& +35:54:52.6 &   14.50 &  0.24 &   18.32 & 0.10 & ... & ... \\
251 & GMRT~J202131.1+354706 & 20:21:31.19& +35:47:06.7 &    8.45 &  0.21 &    8.04 & 0.07 & ... & ... \\
252 & GMRT~J202131.2+354338 & 20:21:31.24& +35:43:38.1 &    3.86 &  0.17 &    3.56 & 0.06 & ... & ... \\
253 & GMRT~J202133.8+355005 & 20:21:33.81& +35:50:05.7 &  209.34 &  0.27 &  240.10 & 0.14 & ... & ... \\
254 & GMRT~J202135.6+370950 & 20:21:35.67& +37:09:50.5 &   49.01 &  0.19 &  104.52 & 0.13 & ND  & ND  \\
255 & GMRT~J202136.0+382122 & 20:21:36.05& +38:21:22.0 &    3.12 &  0.23 &    2.59 & 0.07 & ... & ... \\
256 & GMRT~J202136.2+380704 & 20:21:36.26& +38:07:04.3 &    3.63 &  0.19 &    5.10 & 0.08 & ... & ... \\
257 & GMRT~J202136.7+381812 & 20:21:36.71& +38:18:12.0 &    3.40 &  0.24 &    4.69 & 0.09 & ... & ... \\
258 & GMRT~J202138.4+373110 & 20:21:38.48& +37:31:10.8 &   68.13 &  0.32 & 3090.77 & 0.90 & ... & ... \\
259 & GMRT~J202141.3+372557 & 20:21:41.36& +37:25:57.6 &   54.13 &  0.53 & 2866.42 & 1.50 & ... & ... \\
260 & GMRT~J202143.3+353334 & 20:21:43.34& +35:33:34.1 &    3.01 &  0.17 &    4.06 & 0.07 & ... & ... \\
261 & GMRT~J202147.7+363929 & 20:21:47.76& +36:39:29.2 &   37.14 &  0.24 &   48.73 & 0.11 & ND  & ... \\
262 & GMRT~J202149.1+373301 & 20:21:49.13& +37:33:01.6 &    6.15 &  0.30 &   11.48 & 0.18 & ... & ... \\
263 & GMRT~J202149.5+362526 & 20:21:49.55& +36:25:26.9 &    8.66 &  0.21 &    9.71 & 0.08 & ND  & ... \\
264 & GMRT~J202149.6+364323 & 20:21:49.60& +36:43:23.0 &   71.92 &  0.29 &  175.73 & 0.18 & ND  & ... \\
265 & GMRT~J202150.4+373014 & 20:21:50.40& +37:30:14.2 &    9.00 &  0.32 &   48.84 & 0.40 & ... & ... \\
266 & GMRT~J202153.8+355622 & 20:21:53.84& +35:56:22.7 &   18.53 &  0.23 &   27.50 & 0.11 & ... & ND  \\
267 & GMRT~J202154.4+354339 & 20:21:54.41& +35:43:39.3 &    5.05 &  0.17 &    4.56 & 0.06 & ... & ... \\
268 & GMRT~J202154.6+374619 & 20:21:54.62& +37:46:19.2 &   48.56 &  0.19 &   56.53 & 0.08 & ... & ... \\
269 & GMRT~J202158.0+370938 & 20:21:58.06& +37:09:38.7 &    2.69 &  0.23 &    4.51 & 0.09 & ... & ... \\
270 & GMRT~J202158.4+380651 & 20:21:58.44& +38:06:51.6 &    2.52 &  0.19 &    3.01 & 0.07 & ... & ... \\
271 & GMRT~J202158.6+354335 & 20:21:58.66& +35:43:35.7 &   15.72 &  0.17 &   27.36 & 0.09 & ... & ... \\
272 & GMRT~J202201.3+361112 & 20:22:01.38& +36:11:12.1 &   51.45 &  0.24 &   78.39 & 0.13 & ... & ... \\
273 & GMRT~J202207.8+373006 & 20:22:07.89& +37:30:06.9 &    2.28 &  0.27 &    5.16 & 0.10 & ... & ... \\
274 & GMRT~J202209.9+352927 & 20:22:09.98& +35:29:27.6 &   11.60 &  0.19 &   13.90 & 0.08 & ... & ... \\
275 & GMRT~J202210.7+380826 & 20:22:10.78& +38:08:26.1 &    5.83 &  0.22 &    8.16 & 0.11 & ... & ... \\
276 & GMRT~J202213.2+355940 & 20:22:13.21& +35:59:40.8 &    4.23 &  0.23 &    5.52 & 0.09 & ... & ... \\
277 & GMRT~J202214.1+370031 & 20:22:14.10& +37:00:31.4 &    3.31 &  0.23 &    3.44 & 0.07 & ... & ... \\
278 & GMRT~J202216.7+373059 & 20:22:16.75& +37:30:59.5 &    4.62 &  0.24 &    5.20 & 0.11 & ... & ... \\
279 & GMRT~J202218.7+365421 & 20:22:18.71& +36:54:21.6 &   35.30 &  0.28 &   56.32 & 0.13 & ... & ... \\
280 & GMRT~J202218.8+362526 & 20:22:18.86& +36:25:26.6 &    2.92 &  0.24 &    3.17 & 0.07 & ... & ... \\
281 & GMRT~J202220.5+374759 & 20:22:20.59& +37:47:59.5 &   30.61 &  0.15 &   32.57 & 0.07 & ... & ... \\
282 & GMRT~J202225.1+363138 & 20:22:25.12& +36:31:38.2 &    3.73 &  0.25 &    4.12 & 0.09 & ... & ... \\
283 & GMRT~J202225.8+355521 & 20:22:25.83& +35:55:21.7 &   38.20 &  0.22 &   43.09 & 0.09 & ... & ... \\
284 & GMRT~J202230.0+370105 & 20:22:30.00& +37:01:05.8 &    5.17 &  0.24 &    9.57 & 0.11 & ... & ... \\
285 & GMRT~J202231.6+375537 & 20:22:31.64& +37:55:37.1 &   17.90 &  0.16 &   24.74 & 0.07 & ... & ... \\
286 & GMRT~J202235.5+351935 & 20:22:35.59& +35:19:35.6 &   57.51 &  0.33 &   87.06 & 0.16 & ... & ... \\
287 & GMRT~J202236.1+351940 & 20:22:36.18& +35:19:40.5 &   14.35 &  0.33 &   22.32 & 0.15 & ... & ... \\
288 & GMRT~J202242.1+380342 & 20:22:42.14& +38:03:42.1 &    5.23 &  0.17 &    5.70 & 0.06 & ... & ... \\
289 & GMRT~J202242.7+353445 & 20:22:42.73& +35:34:45.3 &    3.44 &  0.18 &    3.43 & 0.06 & ... & ... \\
290 & GMRT~J202242.9+373227 & 20:22:42.91& +37:32:27.9 &    3.72 &  0.23 &    4.11 & 0.08 & ... & ... \\
291 & GMRT~J202248.9+373927 & 20:22:48.94& +37:39:27.7 &    8.79 &  0.17 &    8.70 & 0.06 & ... & ... \\
292 & GMRT~J202251.6+353326 & 20:22:51.67& +35:33:26.4 &    5.02 &  0.18 &    4.89 & 0.06 & ... & ... \\
293 & GMRT~J202258.2+361625 & 20:22:58.28& +36:16:25.2 &    7.58 &  0.19 &   12.72 & 0.11 & ... & ... \\
294 & GMRT~J202258.5+361147 & 20:22:58.50& +36:11:47.2 &    4.62 &  0.23 &    4.86 & 0.08 & ... & ... \\
295 & GMRT~J202259.9+375020 & 20:22:59.95& +37:50:20.4 &   24.07 &  0.17 &   24.95 & 0.07 & ... & ... \\
296 & GMRT~J202303.7+352129 & 20:23:03.73& +35:21:29.3 &   99.10 &  0.37 &  145.95 & 0.18 & ... & ... \\
297 & GMRT~J202308.0+355829 & 20:23:08.00& +35:58:29.5 &    4.90 &  0.35 &    5.98 & 0.12 & ... & ... \\
298 & GMRT~J202309.0+373641 & 20:23:09.09& +37:36:41.1 &    4.14 &  0.17 &    4.15 & 0.07 & ... & ... \\
299 & GMRT~J202313.5+374833 & 20:23:13.57& +37:48:33.5 &   95.36 &  0.20 &  133.27 & 0.10 & ... & ... \\
300 & GMRT~J202319.3+351811 & 20:23:19.35& +35:18:11.5 &   26.57 &  0.46 &   38.88 & 0.20 & ... & ... \\
301 & GMRT~J202320.8+362834 & 20:23:20.83& +36:28:34.3 &    6.87 &  0.20 &   18.72 & 0.15 & ... & ... \\
302 & GMRT~J202329.2+375111 & 20:23:29.29& +37:51:11.3 &    4.45 &  0.16 &    4.98 & 0.06 & ... & ... \\
303 & GMRT~J202329.9+371530 & 20:23:29.96& +37:15:30.1 &    9.06 &  0.19 &   10.03 & 0.09 & ... & ... \\
304 & GMRT~J202339.9+355100 & 20:23:39.94& +35:51:00.2 &   55.50 &  0.33 &   79.15 & 0.15 & ... & ... \\
305 & GMRT~J202345.3+372731 & 20:23:45.35& +37:27:31.8 &   13.46 &  0.23 &   19.61 & 0.10 & ... & ... \\
306 & GMRT~J202347.8+373425 & 20:23:47.83& +37:34:25.4 &    8.91 &  0.21 &   10.22 & 0.08 & ... & ... \\
307 & GMRT~J202351.4+375343 & 20:23:51.44& +37:53:43.9 &    4.04 &  0.16 &    3.78 & 0.06 & ... & ... \\
308 & GMRT~J202355.0+373810 & 20:23:55.03& +37:38:10.0 &   50.73 &  0.22 &  249.27 & 0.18 & ... & ... \\
309 & GMRT~J202403.3+360803 & 20:24:03.38& +36:08:03.8 &    4.32 &  0.30 &    7.94 & 0.14 & ... & ... \\
310 & GMRT~J202404.2+380216 & 20:24:04.20& +38:02:16.4 &    8.05 &  0.21 &    7.21 & 0.07 & ... & ... \\
311 & GMRT~J202404.2+374300 & 20:24:04.29& +37:43:00.6 &   30.30 &  0.18 &   93.28 & 0.13 & ... & ... \\
312 & GMRT~J202407.1+360032 & 20:24:07.17& +36:00:32.7 &    9.59 &  0.42 &   12.26 & 0.18 & ... & ... \\
313 & GMRT~J202415.5+360821 & 20:24:15.57& +36:08:21.1 &    6.93 &  0.30 &    8.63 & 0.13 & ... & ... \\
314 & GMRT~J202420.6+355043 & 20:24:20.60& +35:50:43.6 &    3.76 &  0.34 &    7.61 & 0.13 & ... & ... \\
315 & GMRT~J202422.3+365518 & 20:24:22.37& +36:55:18.0 &   45.56 &  0.27 &   62.30 & 0.12 & ... & ... \\
316 & GMRT~J202422.6+355354 & 20:24:22.60& +35:53:54.8 &    9.21 &  0.32 &    9.06 & 0.11 & ... & ... \\
317 & GMRT~J202424.6+373354 & 20:24:24.67& +37:33:54.3 &   10.12 &  0.20 &   13.81 & 0.10 & ... & ... \\
318 & GMRT~J202430.6+364604 & 20:24:30.60& +36:46:04.1 &   20.59 &  0.25 &   50.61 & 0.16 & ... & ... \\
319 & GMRT~J202431.0+365139 & 20:24:31.00& +36:51:39.0 &   10.79 &  0.26 &   16.51 & 0.11 & ... & ... \\
320 & GMRT~J202431.7+363020 & 20:24:31.73& +36:30:20.0 &   19.59 &  0.36 &   35.91 & 0.21 & ... & ... \\
321 & GMRT~J202432.0+372355 & 20:24:32.00& +37:23:55.0 &   11.04 &  0.24 &   15.61 & 0.11 & ... & ... \\
322 & GMRT~J202432.0+363012 & 20:24:32.03& +36:30:12.1 &   74.10 &  0.36 &  121.98 & 0.20 & ... & ... \\
323 & GMRT~J202446.1+370236 & 20:24:46.12& +37:02:36.2 &    4.04 &  0.28 &    8.03 & 0.13 & ... & ... \\
324 & GMRT~J202450.0+361706 & 20:24:50.02& +36:17:06.2 &    4.76 &  0.39 &    4.42 & 0.11 & ... & ... \\
325 & GMRT~J202450.7+370300 & 20:24:50.74& +37:03:00.8 &   16.17 &  0.28 &   40.96 & 0.17 & ... & ... \\
326 & GMRT~J202454.1+361501 & 20:24:54.13& +36:15:01.6 &   20.39 &  0.41 &   40.35 & 0.20 & ... & ... \\
327 & GMRT~J202455.3+375744 & 20:24:55.33& +37:57:44.8 &   10.70 &  0.33 &   12.62 & 0.11 & ... & ... \\
328 & GMRT~J202504.3+370312 & 20:25:04.35& +37:03:12.4 &    4.08 &  0.27 &    6.65 & 0.12 & ... & ... \\
329 & GMRT~J202505.1+370256 & 20:25:05.18& +37:02:56.0 &    4.02 &  0.26 &    5.06 & 0.10 & ... & ... \\
330 & GMRT~J202508.4+362535 & 20:25:08.42& +36:25:35.4 &    8.16 &  0.46 &   10.36 & 0.17 & ... & ... \\
331 & GMRT~J202509.4+353642 & 20:25:09.45& +35:36:42.5 &    8.96 &  0.52 &   47.92 & 0.50 & ... & ... \\
332 & GMRT~J202515.9+370916 & 20:25:15.91& +37:09:16.8 &   22.68 &  0.28 &   25.20 & 0.11 & ... & ... \\
333 & GMRT~J202523.5+372901 & 20:25:23.59& +37:29:01.5 &   43.45 &  0.18 &   48.67 & 0.08 & ... & ... \\
334 & GMRT~J202523.6+372314 & 20:25:23.61& +37:23:14.8 &   34.09 &  0.17 & 1033.75 & 0.50 & ... & ... \\
335 & GMRT~J202532.1+354459 & 20:25:32.14& +35:44:59.8 &    4.59 &  0.32 &    5.35 & 0.12 & ... & ... \\
336 & GMRT~J202538.5+372024 & 20:25:38.50& +37:20:24.9 &    2.69 &  0.14 &    2.88 & 0.06 & ... & ... \\
337 & GMRT~J202542.3+374942 & 20:25:42.36& +37:49:42.4 &    5.18 &  0.42 &    6.55 & 0.13 & ... & ... \\
338 & GMRT~J202543.5+370607 & 20:25:43.57& +37:06:07.2 &    6.42 &  0.26 &    6.32 & 0.08 & ... & ... \\
339 & GMRT~J202543.6+374917 & 20:25:43.67& +37:49:17.5 &   10.20 &  0.40 &   15.01 & 0.15 & ... & ... \\
340 & GMRT~J202555.7+361553 & 20:25:55.73& +36:15:53.9 &    6.17 &  0.42 &    7.03 & 0.16 & ... & ... \\
341 & GMRT~J202556.3+365011 & 20:25:56.38& +36:50:11.4 &    5.15 &  0.21 &    4.42 & 0.06 & ... & ... \\
342 & GMRT~J202603.5+363628 & 20:26:03.59& +36:36:28.7 &   24.43 &  0.26 &   25.51 & 0.10 & ... & ... \\
343 & GMRT~J202605.9+361108 & 20:26:05.97& +36:11:08.3 &   10.66 &  0.43 &   10.25 & 0.12 & ... & ... \\
344 & GMRT~J202608.3+360111 & 20:26:08.38& +36:01:11.1 &    7.62 &  0.32 &   12.67 & 0.15 & ... & ... \\
345 & GMRT~J202611.1+372845 & 20:26:11.14& +37:28:45.6 &    7.54 &  0.16 &   10.10 & 0.10 & ... & ... \\
346 & GMRT~J202616.7+360053 & 20:26:16.72& +36:00:53.2 &    5.42 &  0.34 &    7.58 & 0.14 & ... & ... \\
347 & GMRT~J202625.8+365929 & 20:26:25.85& +36:59:29.6 &    6.84 &  0.27 &   10.33 & 0.13 & ... & ND  \\
348 & GMRT~J202625.9+365319 & 20:26:25.90& +36:53:19.1 &    5.64 &  0.25 &    5.82 & 0.11 & ... & $1.9\pm0.4\times10^{-14}$ \\
349 & GMRT~J202626.8+363712 & 20:26:26.87& +36:37:12.6 &    5.32 &  0.25 &    5.82 & 0.09 & ... & ... \\
350 & GMRT~J202629.6+370513 & 20:26:29.65& +37:05:13.6 &    5.56 &  0.29 &    5.94 & 0.09 & ... & ... \\
351 & GMRT~J202632.8+371147 & 20:26:32.80& +37:11:47.6 &    2.54 &  0.20 &    2.52 & 0.06 & ... & ... \\
352 & GMRT~J202638.4+370730 & 20:26:38.42& +37:07:30.1 &    9.79 &  0.24 &    6.07 & 0.08 & ... & ... \\
353 & GMRT~J202638.8+370728 & 20:26:38.84& +37:07:28.1 &   16.74 &  0.30 &   41.84 & 0.20 & ... & ... \\
354 & GMRT~J202645.0+370022 & 20:26:45.04& +37:00:22.1 &   42.16 &  0.32 &   57.57 & 0.14 & ... & ... \\
355 & GMRT~J202647.2+370613 & 20:26:47.23& +37:06:13.3 &    4.88 &  0.32 &    9.50 & 0.16 & ... & ... \\
356 & GMRT~J202703.3+374853 & 20:27:03.33& +37:48:53.5 &   70.63 &  0.48 &   99.53 & 0.21 & ... & ... \\
357 & GMRT~J202712.4+365818 & 20:27:12.44& +36:58:18.8 &    8.85 &  0.48 &   10.46 & 0.19 & ... & ... \\
358 & GMRT~J202724.8+371042 & 20:27:24.88& +37:10:42.6 &   48.80 &  0.38 &   89.73 & 0.20 & ... & ND  \\
359 & GMRT~J202727.3+372258 & 20:27:27.32& +37:22:58.4 &   72.93 &  0.50 & 5013.70 & 1.40 & ... & ND  \\
360 & GMRT~J202730.3+371523 & 20:27:30.31& +37:15:23.2 &   10.13 &  0.32 &   14.10 & 0.17 & ... & ... \\
361 & GMRT~J202733.3+373116 & 20:27:33.32& +37:31:16.7 &    4.32 &  0.30 &   25.96 & 0.30 & ... & ... \\
362 & GMRT~J202735.0+373124 & 20:27:35.00& +37:31:24.6 &    5.64 &  0.31 &   13.32 & 0.18 & ... & ... \\
\hline
\label{table:list}
\end{longtable}
}

}

\end{document}